\documentstyle[galley,graphicx,epsf,referee]{mn} 
\begin{document}
\title{Optical photometry and basic parameters of 10 unstudied open clusters}
\author[Subramaniam et al.]        
{ Annapurni Subramaniam$^{1}$, 
Giovanni Carraro$^{2}$, 
and Kenneth A. Janes$^3$
\thanks{email: 
purni@iiap.res.in(AS), gcarraro@eso.org(GC), janes@bu.edu(KAJ)}\\ 
$^1$,Indian Institute of Astrophysics, II Block Koramangala, 
Bangalore 560034, India \\
$^2$ European Southern Observatory, Alonso de Cordova 3107
Vitacura, Santiago de Chile, Chile\\
$^3$Department of Astronomy, Boston University, 725 Commonwealth Avenue, Boston, MA 02215,
USA\\
} 
 
\date{\it Submitted: October 2009} 
\pubyear{2010} 
%
%
\maketitle 
\label{firstpage}
            
\begin{abstract} 
We present $BVI$ CCD photometry of 
10 northern open clusters, 
Berkeley 43, Berkeley 45, Berkeley 47, NGC 6846,
Berkeley 49, Berkeley 51,
Berkeley 89, Berkeley 91, Tombaugh 4 and Berkeley 9, and
estimate their fundamental parameters. Eight of the clusters are located in the
first galactic quadrant and 2 are in the second. 
This is the first optical photometry for
8 clusters. All of them  are embedded in rich galactic fields and have large
reddening towards them (E(B$-$V) = 1.0 -- 2.3 mag).  There is a possibility
that some of these difficult-to-study clusters may be asterisms rather than
physical systems, but assuming they are physical clusters, 
we find that 8 of them are located
beyond 2 kpc, and 6 clusters (60\% of the sample) are located well above or 
below the Galactic plane.
Seven clusters have ages 500 Myr or less and the other 3 are 1 Gyr or more in
age.
This sample of clusters has increased the optical photometry of clusters 
in the second half of the first galactic quadrant, beyond 2 kpc, 
from 10 to 15. NGC 6846 is found to be one of the
most distant clusters in this region of the Galaxy.
 
\end{abstract}
 
\section{Introduction} 
Open star clusters are important constituents of our Galactic disk.
They are potential tracers of the structure, star formation
history and the chemical evolution of the disk.
One of the recent such attempts to increase the cluster sample
in the first half of the second galactic quadrant (Subramaniam \& Bhatt 2007)
found an outward extension of the Perseus arm consisting of clusters
older than 100 Myr. They also found a mild warp in the Galactic disk, beyond
2 kpc. An earlier study to identify northern candidate old open clusters in
this region was done by Carraro et al. (2006).

With the goal of increasing our knowledge of the cluster system in the first quadrant of
the galaxy,
we present a $BVI$ photometric study of 10 open clusters, 
Berkeley 43, Berkeley 45, Berkeley 47, NGC 6846,
Berkeley 49, Berkeley 51,
Berkeley 89, Berkeley 91, Tombaugh 4 and Berkeley 9.
These clusters are embedded in rich Galactic fields so little is known about
them; this is the first optical CCD photometric study
for eight of the clusters. 
Berkeley 43 was studied by Hasegawa et al. (2008) and Berkeley 9 was
studied by Maciejewski \& Niedzielski (2007).
Seven of our Berkeley candidates (except Berkeley 9) are in common with the
study by Tadross (2008), using near infared JHK 2MASS 
data. Moffat \& Vogt (1973) obtained UBV photographic data 
of stars in the field of Tombaugh 4, but they could not estimate any
of the parameters since the photometry was not deep enough.
Table 1 lists the coordinates of the clusters in the present study.

\noindent
The plan of the paper is as follows. Sect.~2 describes
the observation strategy and reduction technique.
Sect.~3 deals with the radial density profile and the cluster size.
The results for the 10 clusters are presented in section 4, followed by discussion in section 5.

\renewcommand{\thetable}{1}
\begin{table}
\caption{Basic parameters of the clusters under investigation.
Coordinates are for J2000.0.
}
\begin{tabular}{ccccc}
\hline
\hline
\multicolumn{1}{c}{Name} &
\multicolumn{1}{c}{$RA$}  &
\multicolumn{1}{c}{$DEC$}  &
\multicolumn{1}{c}{$l$} &
\multicolumn{1}{c}{$b$} \\
\hline
& {\rm $hh:mm:ss$} & {\rm $^{o}$~:~$^{\prime}$~:~$^{\prime\prime}$} & [deg] & [deg]\\
\hline
Berkeley 43   & 19:15:36 & +11:13:00 & 45.65 & $-$0.182\\
Berkeley 45   & 19:19:12 & +15:43:00 & 50.04 & 1.145 \\
Berkeley 47   & 19:28:36 & +17:22:06 & 52.561 & $-$0.058\\
NGC 6846      & 19:56:28 & +32:20:54 &  68.69 & 1.92\\
Berkeley 49   & 19:59:31 & +34:38:48 & 70.985 & 2.575\\
Berkeley 51   & 20:24:36 & +46:03:00 & 72.147 & 0.291 \\
Berkeley 89   & 20:24:36 & +46:03:00 & 83.160 & 4.822 \\
Berkeley 91   & 21:10:52 & +48:32:12 & 90.064 & 0.132\\
Tombaugh 4    & 02:28:54 & +61:47:00 &  134.21 & 1.073\\
Berkeley 9    & 03:32:42 & +52:39:00 & 146.07 & -2.82\\
\hline\hline
\end{tabular}
\end{table}

\section{Observations and Data Reduction} 
We obtained our observations with the HFOSC instrument 
at the 2-m Himalayan Chandra Telescope (HCT),
located at Hanle, IAO and operated by Indian Institute of Astrophysics.
Details of the telescope and the instrument are available at the
institute's homepage (http://www.iiap.res.in/).
The CCD used for imaging is a 2 k $\times$ 4 k CCD,
where the central 2 k $\times$ 2 k pixels are used for imaging. The pixel 
size is 15 $\mu$
with an image scale of 0.297 arcsec/pixel. The total area observed is
approximately 10 $\times$ 10 arcmin$^2$. Table 2 is a log of our observations.\\

\noindent
We reduced the data with the
IRAF\footnote{IRAF is distributed by NOAO, which are operated by AURA under
cooperative agreement with the NSF.}
packages CCDRED, DAOPHOT, ALLSTAR and PHOTCAL using the point spread function 
(PSF) method (Stetson 1987).\\

The nights were photometric and we used Landolt (1992)
standard field SA110 for calibration images at different air-masses
during the night to put the photometry onto the standard system.

\renewcommand{\thetable}{2}
\begin{table*}
\centering
\caption{Log of photometric observations}
\begin{tabular}{lrrr}
\hline
Cluster& Date & Filter & Exp time (sec) \\
\hline
NGC 6846 & 30 August 2005 & V & 60, 180, 2X300\\
      &                & B & 60, 120, 300, 600 \\
      &                & I & 10, 30, 3X60\\
Tombaugh 4 & 30 August 2005 & V & 60, 300, 600\\
      &                & B & 120, 2X600 \\
      &                & I & 10, 30, 120, 2X300\\
Be 9  & 31 August 2005 & V & 30, 60, 2X180\\
      &                & B & 60, 300, 600 \\
      &                & I & 10, 30, 2X120\\
Be 43  & 01 August 2008 & V & 30, 120, 180, 240\\
      &                & B &  300, 600\\
      &                & I & 10, 30, 60\\
Be 49  & 01 August 2008 & V & 120, 480\\
      &                & B &  600, 1200\\
      &                & I & 10, 60, 180\\
Be 91  & 01 August 2008 & V & 15, 30, 120\\
      &                & B &  60, 180\\
      &                & I & 5, 20, 60\\
Be 45  & 02 August 2008 & V & 5, 20, 120\\
      &                & B &  300, 600\\
      &                & I & 10, 30, 60\\
Be 47 & 02 August 2008 & V & 10, 60, 120 \\
      &                & B & 60, 240 \\
      &                & I & 0.2, 10, 60 \\
Be 51 & 02 August 2008 & V & 3, 5, 30 \\
      &                & B & 5, 30, 90\\
      &                & I & 2, 10, 60 \\
Be 89 & 02 August 2008 & V & 10, 60, 180 \\
      &                & B & 60, 120, 240 \\
      &                & I & 5, 10, 120\\
\hline
\end{tabular}
\end{table*}

\noindent
The calibration equations are of the form:\\

\noindent
$ b = B + b_1 + b_2 \times X + b_3~(B-V)$ \\
$ v = V + v_1 + v_2 \times X + v_3~(B-V)$ \\
$ i = I + i_1 + i_2 \times X + i_3~(V-I)$ ,\\

\renewcommand{\thetable}{3}
\begin{table*}
\tabcolsep 0.2truecm
\caption {Coefficients of the calibration equations }
\begin{tabular}{ccc}
\hline
30 \& 31 August 2005 & & \\
\hline
$b_1 = 0.818 \pm 0.008$ & $b_2 =  0.26 \pm 0.02$ & $b_3 =  0.043 \pm 0.008$ \\
$v_1 = 0.513 \pm 0.005$ & $v_2 =  0.14 \pm 0.02$ & $v_3 =  0.063 \pm 0.005$ \\
$i_1 = 0.824 \pm 0.009$ & $i_2 =  0.08 \pm 0.02$ & $i_3 =  0.048 \pm 0.009$ \\
\hline
01 August 2008 & & \\
\hline
$b_1 = 1.038 \pm 0.005$ & $b_2 =  0.21 \pm 0.02$ & $b_3 =  0.041 \pm 0.004$ \\
$v_1 = 0.720 \pm 0.008$ & $v_2 =  0.11 \pm 0.02$ & $v_3 =  -0.088 \pm 0.006$ \\
$i_1 = 0.900 \pm 0.010$ & $i_2 =  0.07 \pm 0.02$ & $i_3 =  -0.080 \pm 0.010$ \\
\hline
02 August 2008 & & \\
\hline
$b_1 = 1.020 \pm 0.010$ & $b_2 =  0.20 \pm 0.02$ & $b_3 =  0.052 \pm 0.010$ \\
$v_1 = 0.730 \pm 0.010$ & $v_2 =  0.10 \pm 0.02$ & $v_3 =  -0.079 \pm 0.009$ \\
$i_1 = 0.880 \pm 0.010$ & $i_2 =  0.07 \pm 0.02$ & $i_3 =  -0.050 \pm 0.010$ \\
\hline
\end{tabular}
\end{table*}

\noindent
where $BVI$ are standard magnitudes, $bvi$ are the instrumental ones and  $X$ is
the airmass; all the coefficient values are reported in Table~3.
The standard stars in these fields provide a very good color coverage
(0.1 $\leq (B-V) \leq $ 2.2 and 0.4 $\leq (V-I) \leq $ 2.6)\\

\noindent

We derived aperture correction from a sample of bright stars
and applied them to the photometry.  We found aperture corrections of
0.27, 0.29 and 0.20 mag in B,V and I, respectively.\\

\noindent
The final photometric catalog consisting of 29226 stars present in 10 clusters
(coordinates,
B, V and I magnitudes and errors)
includes 10525 stars in NGC 6846, 5751 in Tombaugh 4, 1461 in Berkeley 9,
1280 in Be 91, 1996 in Be 49, 1141 in Be 43, 2871 in Be 45, 637 in Be 47,
1962 in Be 51 and 1602 in Be 89.  
The limiting magnitudes of the photometry are B = 21.0, V = 22.0 and I =21.0. 
The photometry 
will be made available in electronic form at the
WEBDA\footnote{http://www.univie.ac.at/webda/navigation.html} site
maintained by E. Paunzen.\\

\begin{figure} 
\epsfxsize=9truecm
\epsffile{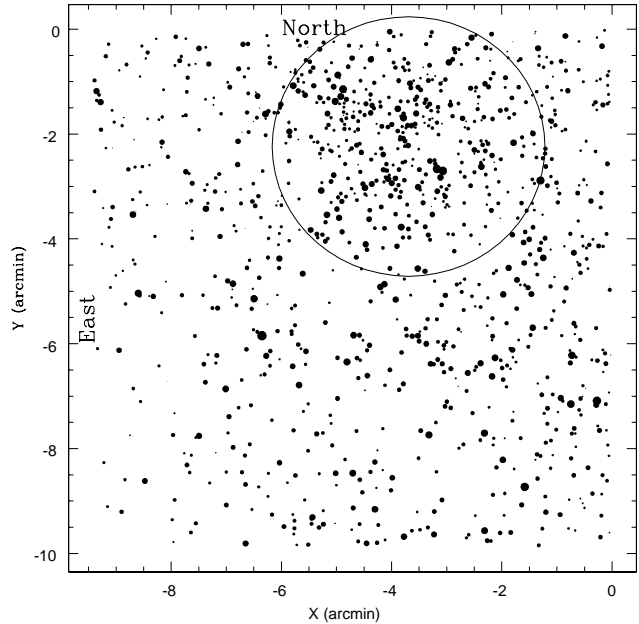}
\caption{The observed region of Berkeley 43. 
We considered stars within the circle of radius of 2.5 arcmin as
shown to be within the effective radius of the cluster.
}
\end{figure} 
\begin{figure} 
\epsfxsize=9truecm
\epsffile{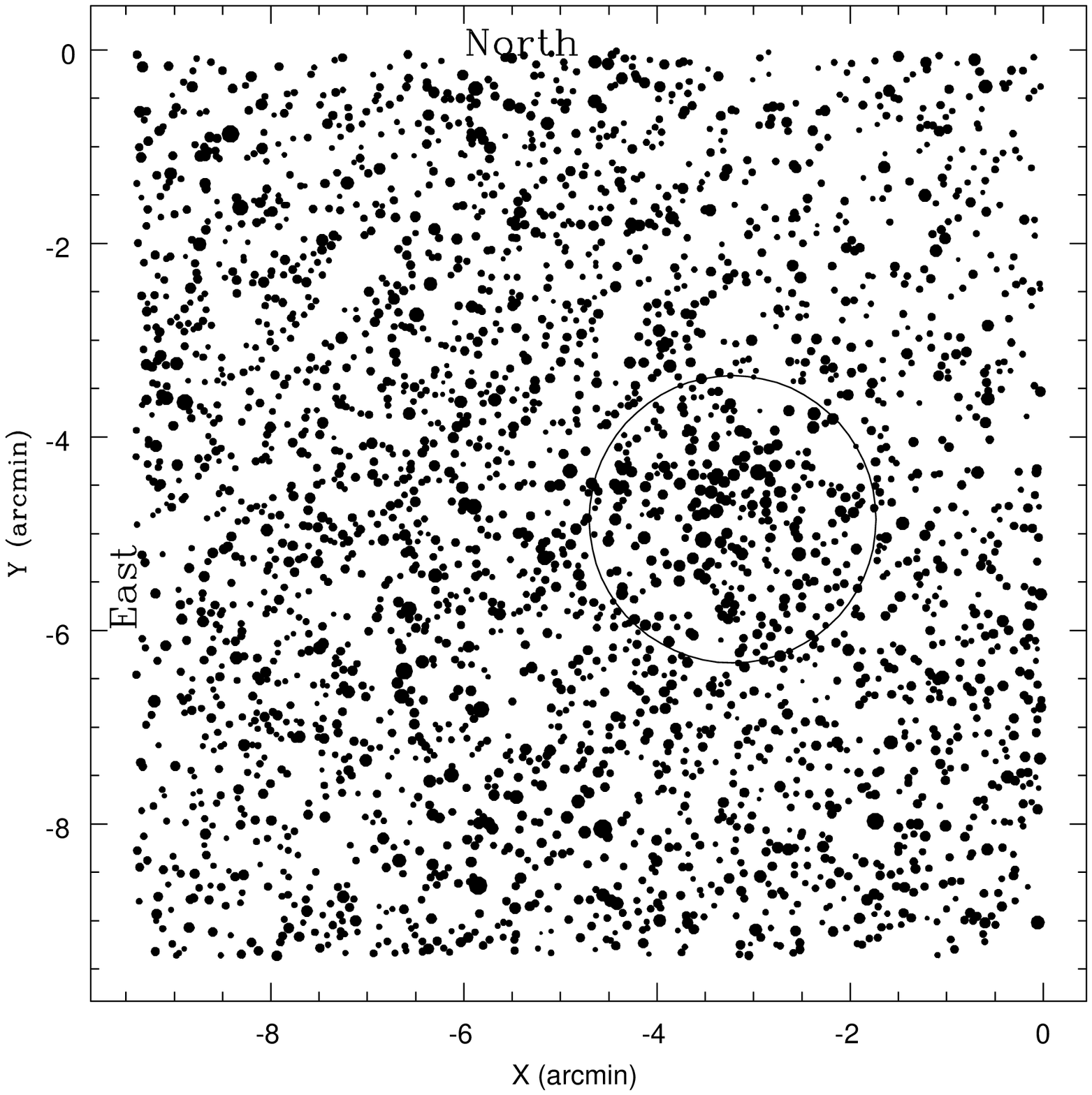}
\caption{The observed region of Berkeley 45.
We considered stars within a radius of 1.5 arcmin as
shown to be within the effective radius of the cluster.
}
\end{figure} 
\begin{figure} 
\epsfxsize=9truecm
\epsffile{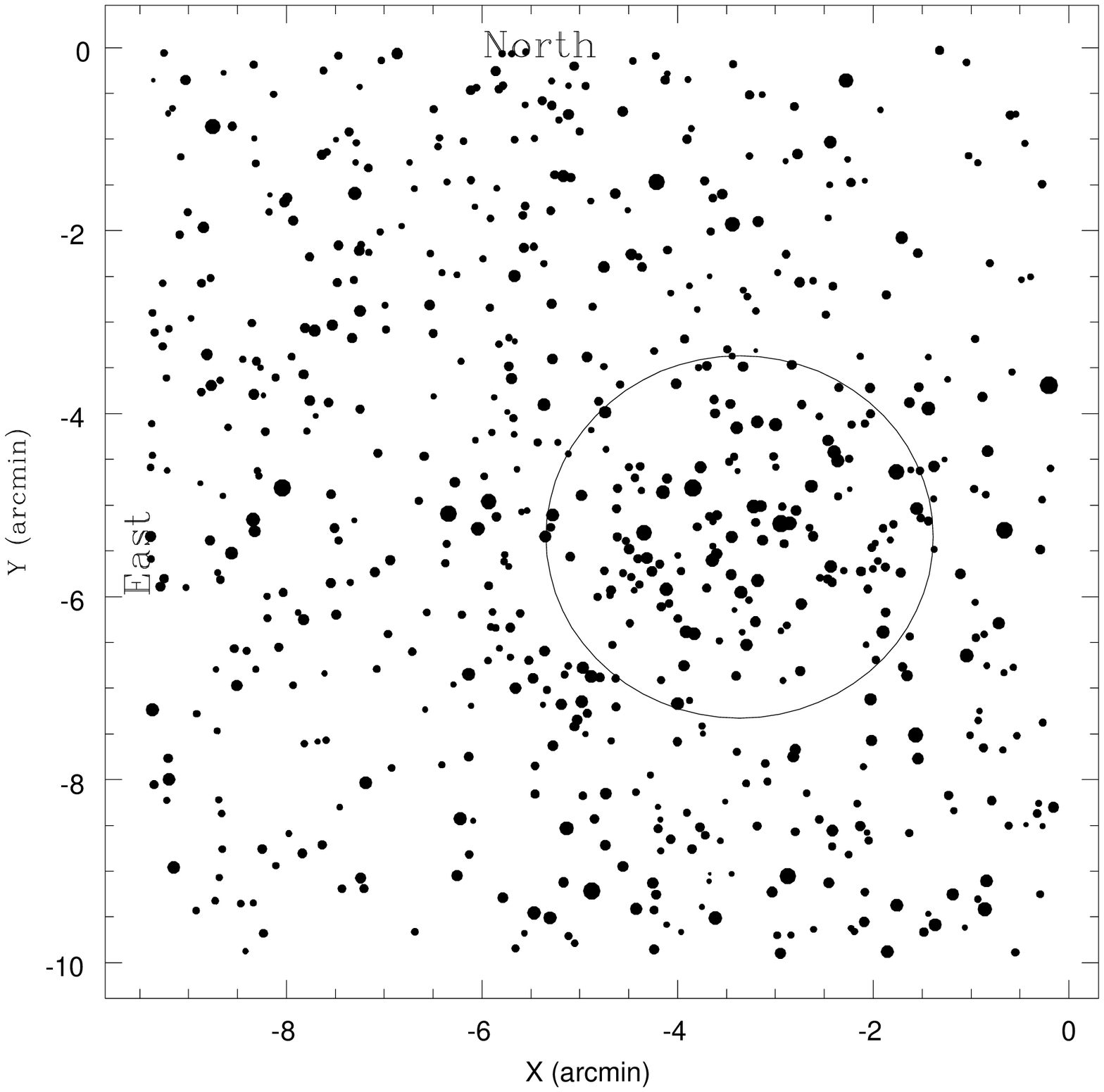}
\caption{The observed region of Berkeley 47. 
We considered stars within a radius of 2.0 arcmin as
shown to be within the effective radius of the cluster.
}
\end{figure} 
\begin{figure} 
\epsfxsize=9truecm
\epsffile{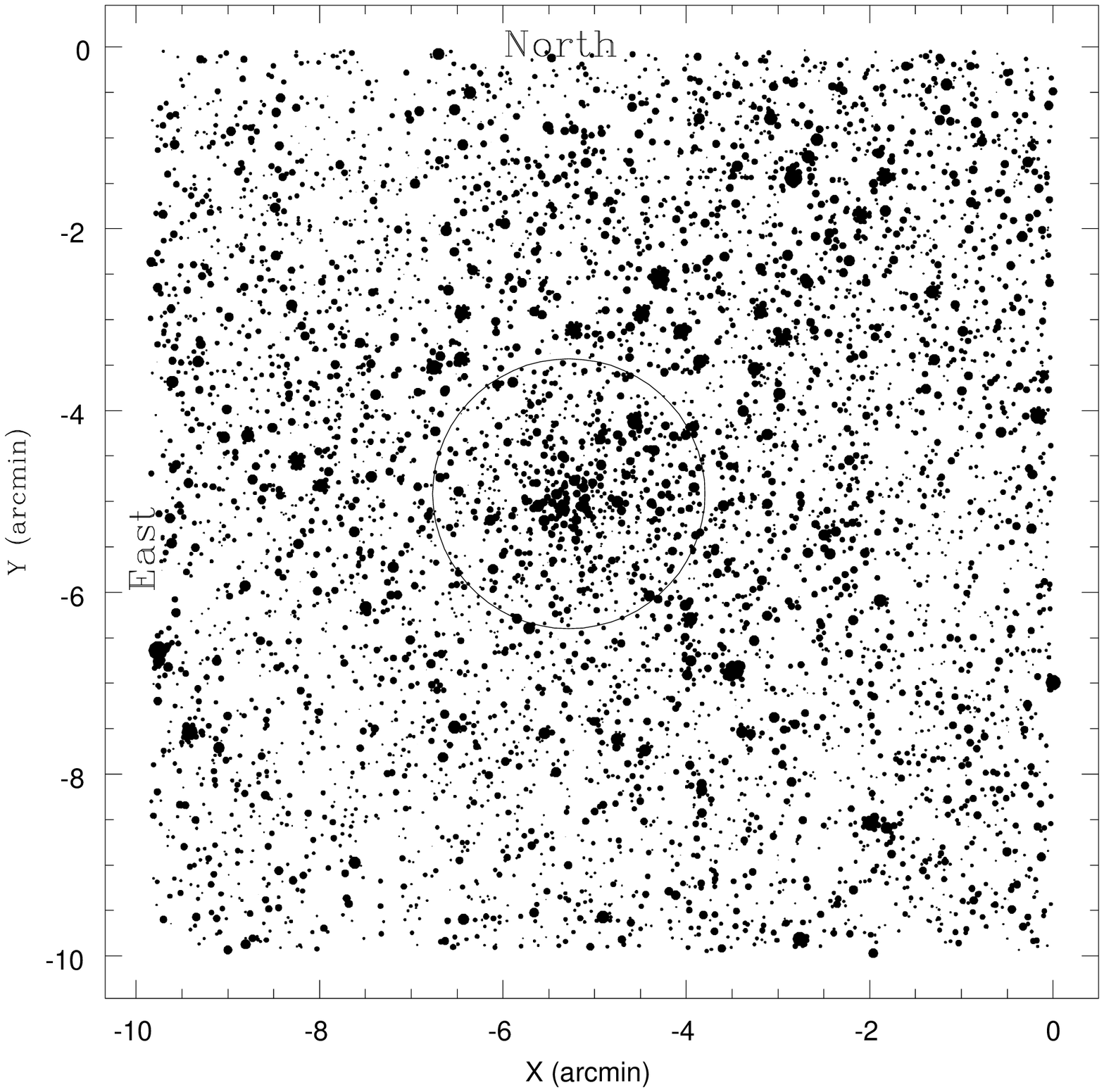} 
\caption{The observed region of NGC 6846. 
We considered stars within a radius of 1.5 arcmin as
shown to be within the effective radius of the cluster.
 }
\end{figure} 

\begin{figure} 
\epsfxsize=9truecm
\epsffile{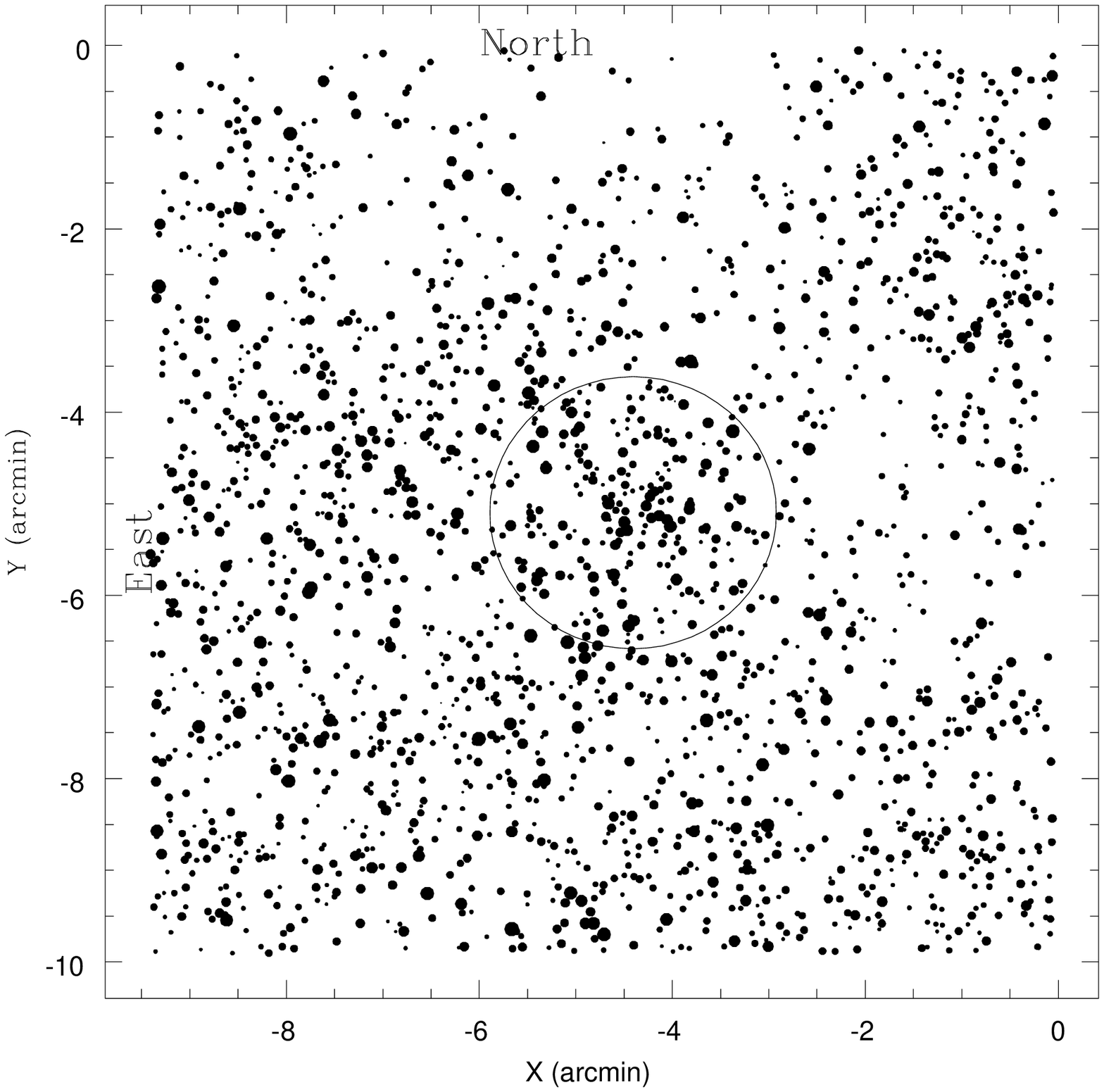}
\caption{The observed region of Berkeley 49.
We considered stars within a radius of 1.5 arcmin as
shown to be within the effective radius of the cluster.
}
\end{figure} 
\begin{figure} 
\epsfxsize=9truecm
\epsffile{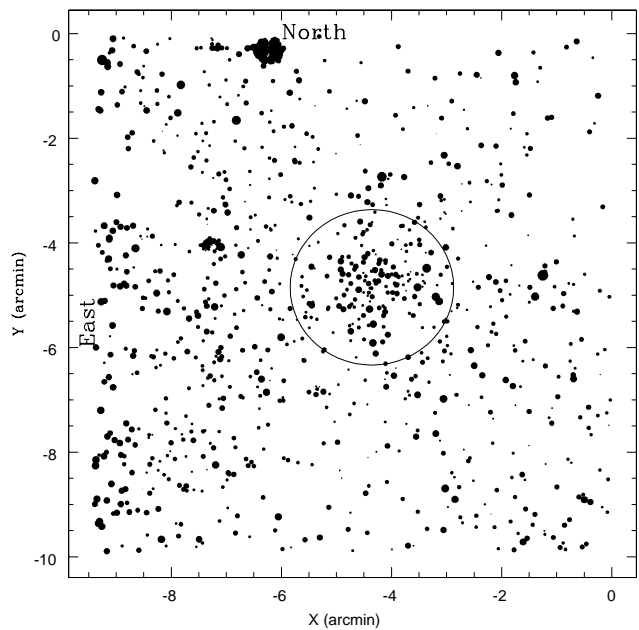}
\caption{The observed region of Berkeley 51. 
Stars within a radius of 1.0 arcmin as
shown are considered to be within the effective radius of the cluster.
}
\end{figure} 
\begin{figure} 
\epsfxsize=9truecm
\epsffile{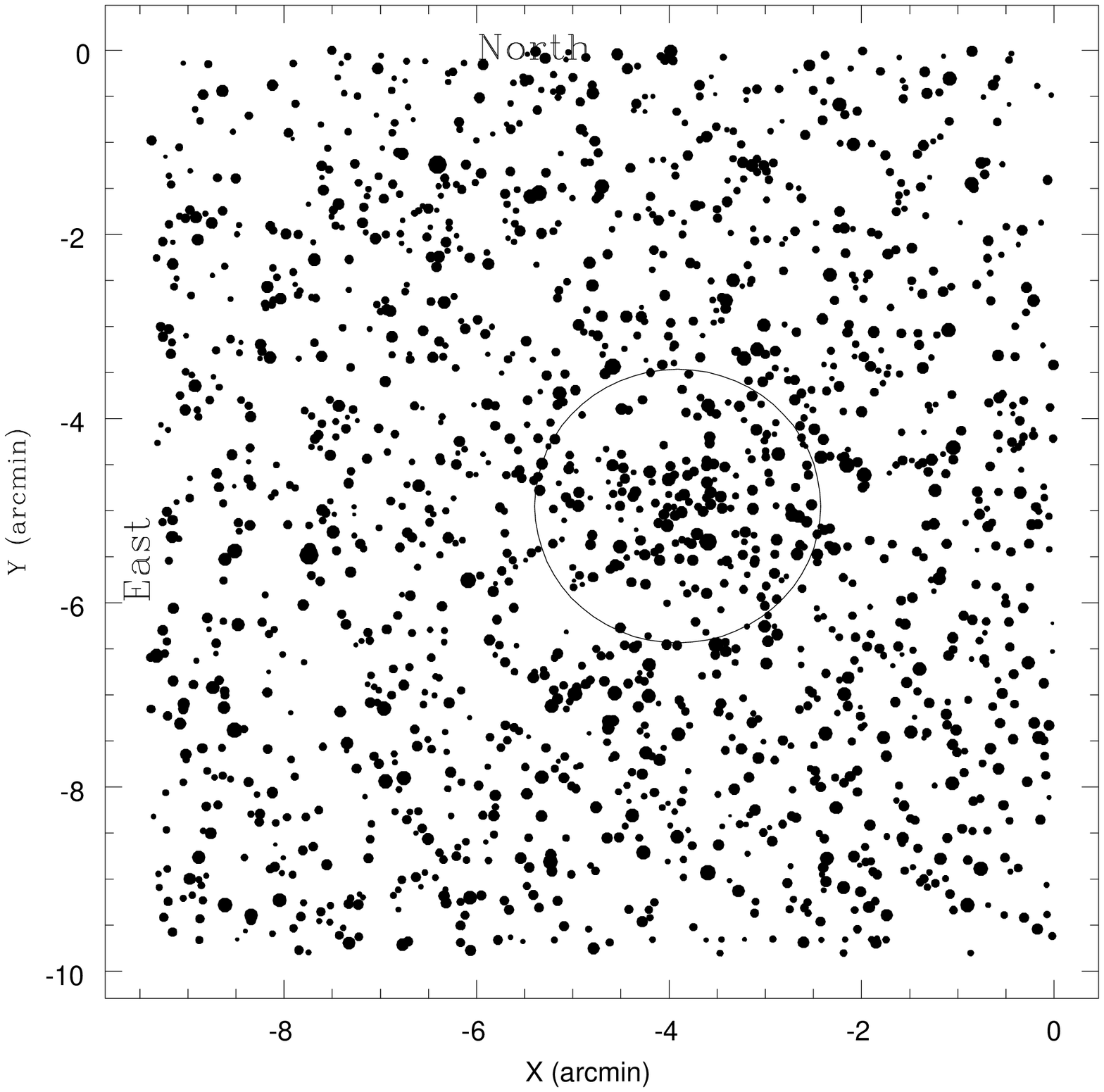}
\caption{The observed region of Berkeley 89.
We considered stars within a radius of 1.0 arcmin as
shown to be within the effective radius of the cluster.
}
\end{figure} 
\begin{figure} 
\epsfxsize=9truecm
\epsffile{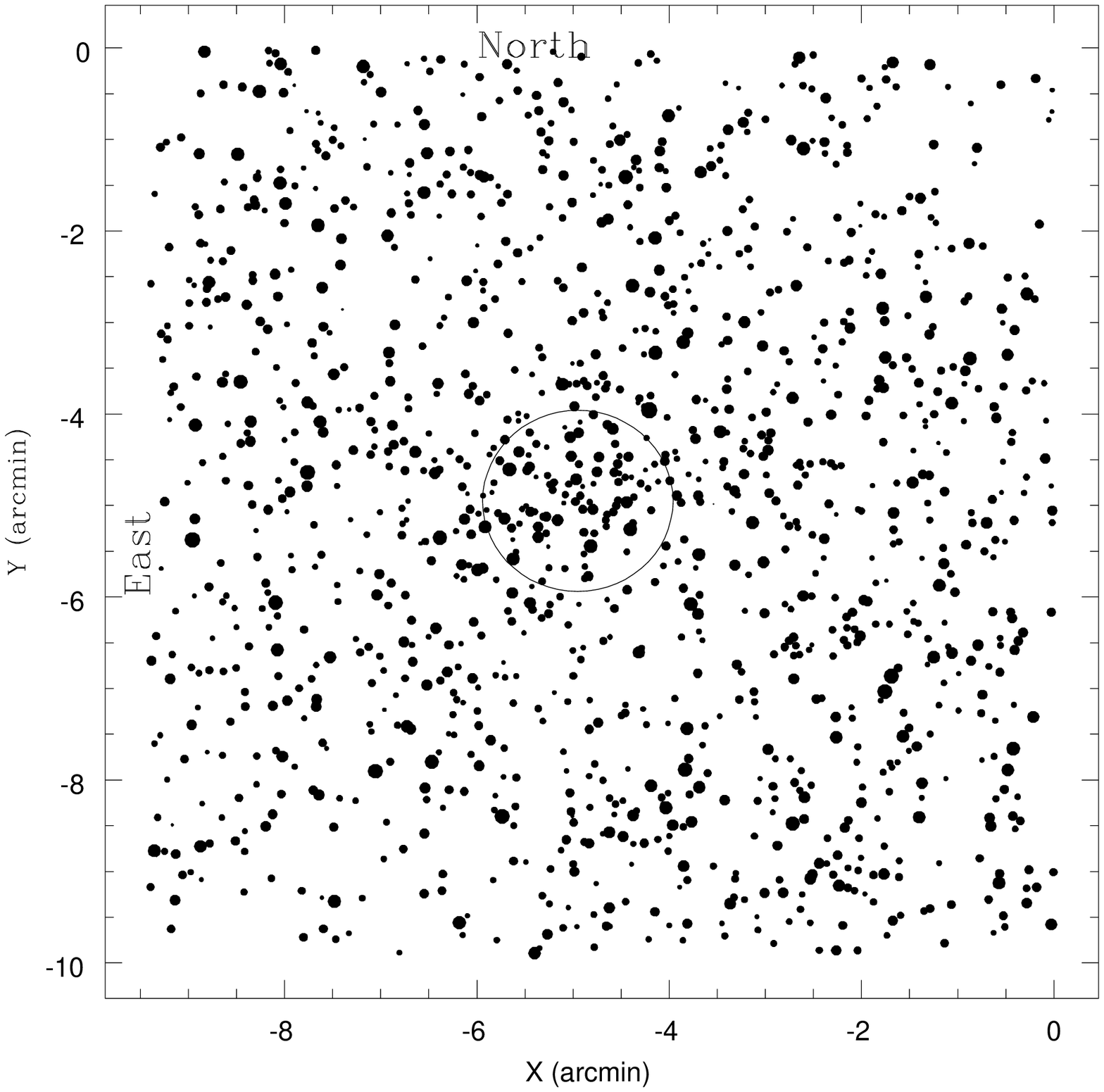}
 \caption{The observed region of Berkeley 91.
We considered stars within a radius of 1.0 arcmin as
shown to be within the effective radius of the cluster.
}
\end{figure} 

\begin{figure} 
\epsfxsize=9truecm
\epsffile{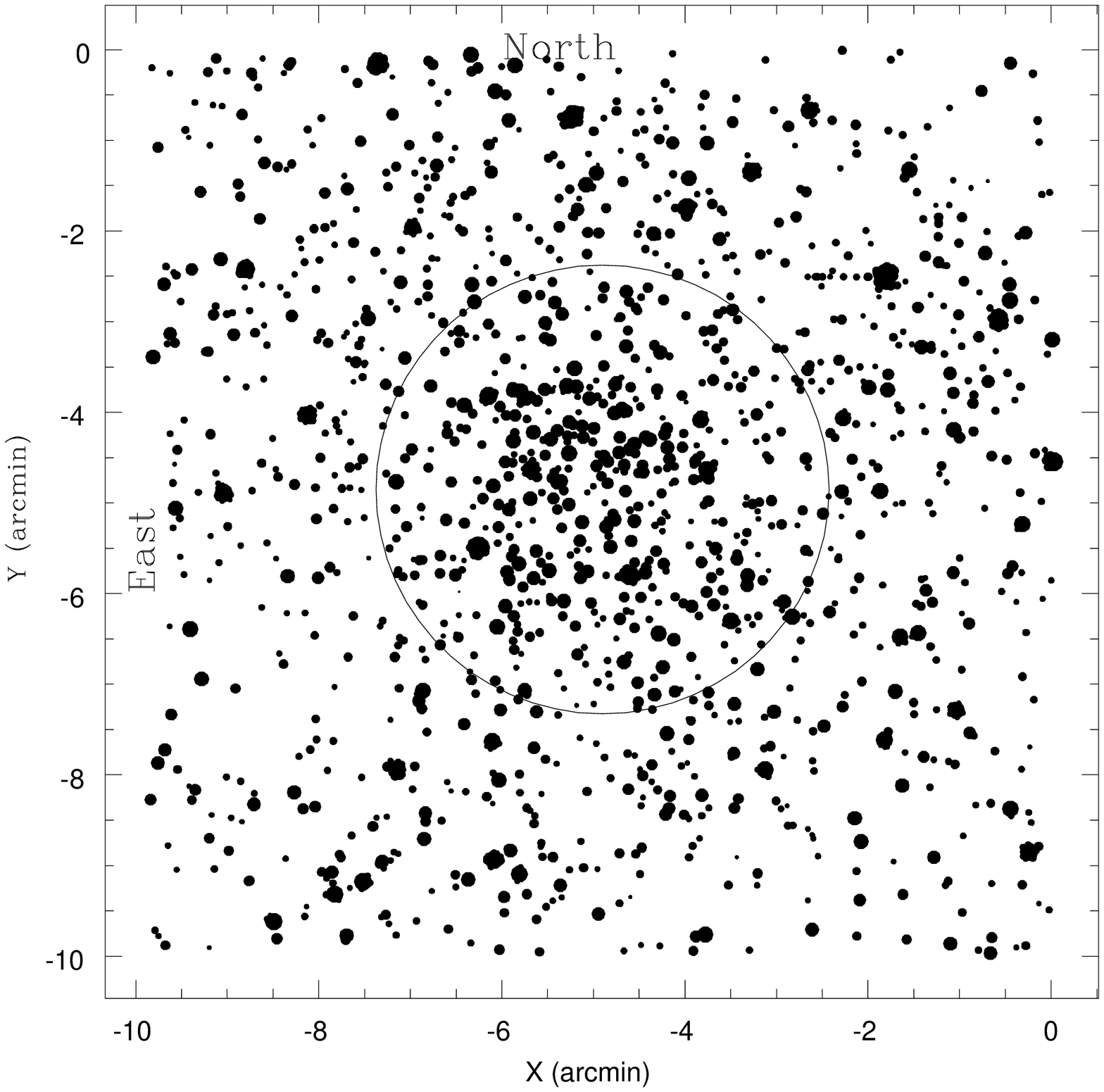} 
\caption{The observed region of Tombaugh 4. 
We considered stars within a radius of 2.5 arcmin as
shown to be within the effective radius of the cluster.
 }
\end{figure} 

\begin{figure} 
\epsfxsize=9truecm
\epsffile{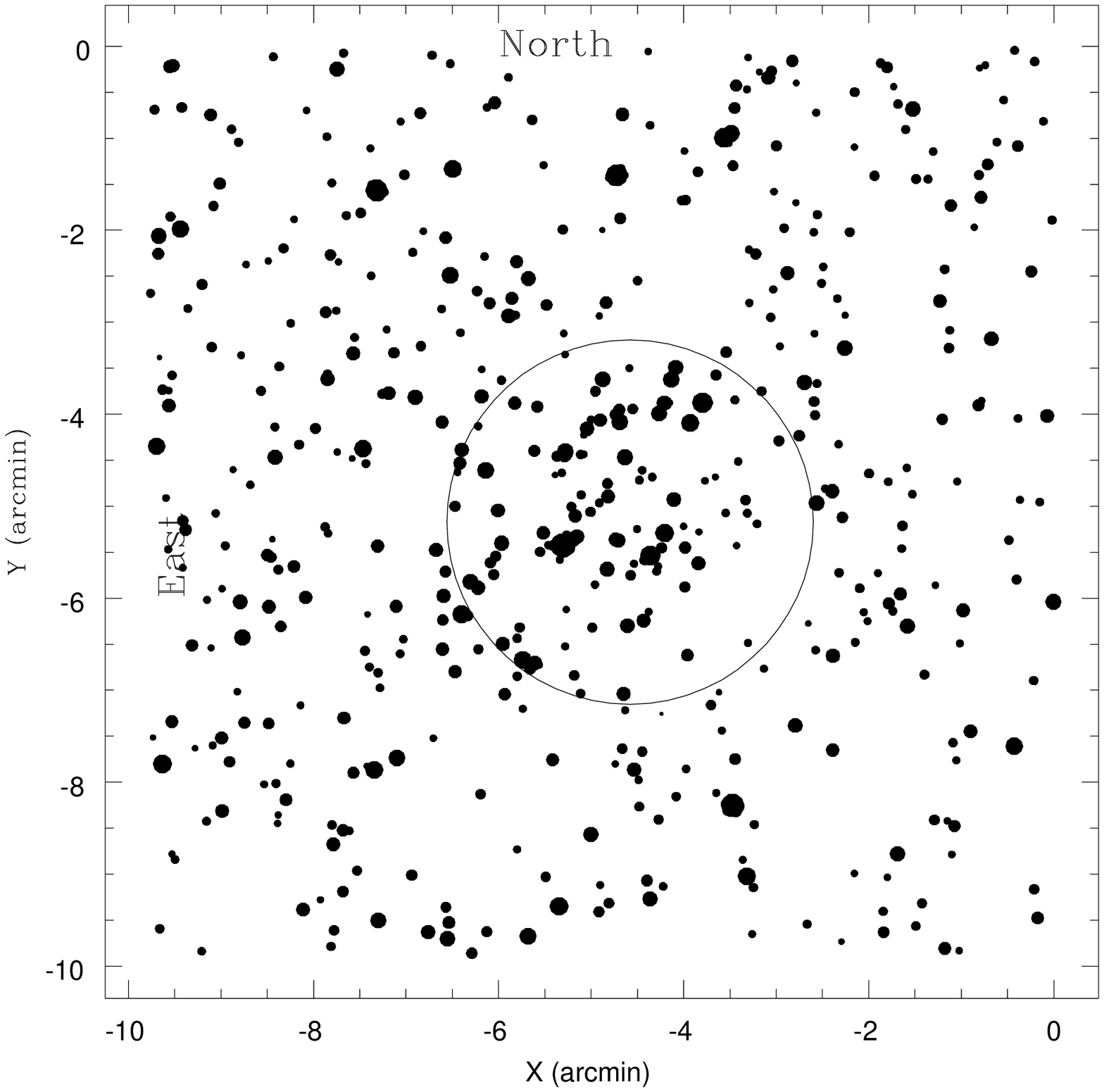}
\caption{The observed region of Berkeley 9. 
We considered stars within a radius of 2.0 arcmin as
shown to be within the effective radius of the cluster.}
\end{figure} 
  
\begin{figure} 
\epsfxsize=9truecm
\epsffile{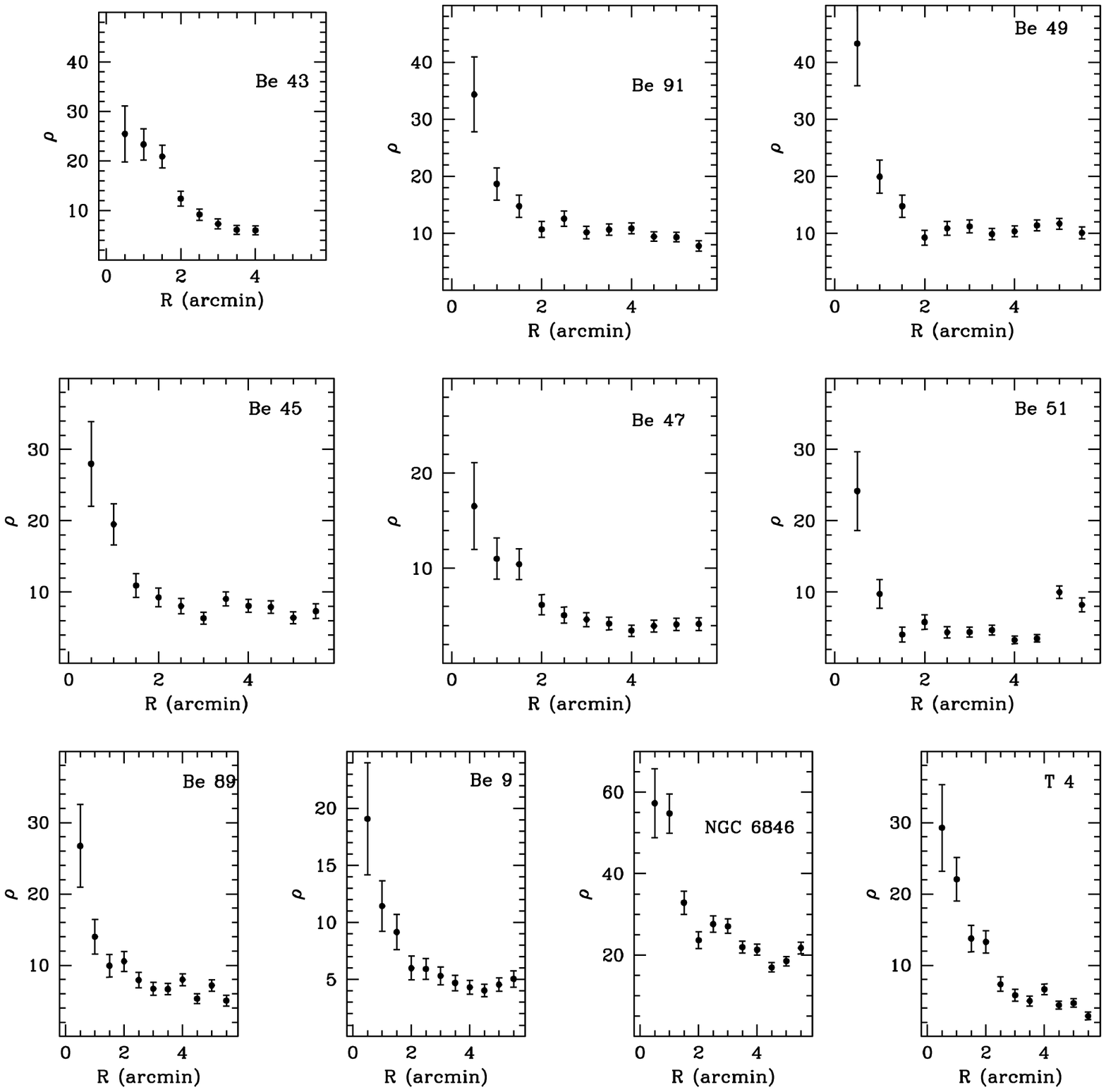} 
\caption{The estimated radial density profiles for 
10 clusters. The cluster names  are indicated inside each figure.}
\end{figure}

\section{Cluster Radii} 
To identify the most likely cluster
members and minimize field star contamination, we derived cluster radii
from radial density profiles (RDP) using
star counts.
We counted stars in circular annuli of 0.5 arcmin wide around
the cluster centers, limiting our counts to stars brighter than V=20 mag.
The stellar density profiles are shown in figure 11.
We visually fitted the RDP with the function
$ \rho(R) \propto f_0/(1+(R/R_0)^2)$, where $R_0$ is the 
radius at which the
density $\rho (R)$ becomes half of the central density, $f_0$.
The estimated half-power radii based on fits to the profiles are 
tabulated in Table 4.

By this measurement, the clusters have radii between 0.6 and 1.9 arcmin, 
making them 
rather small clusters, possibly located relatively far away. In view of
their sparse nature, we simply 
used their visual appearance and the RDP half-power radii as 
guides and rounded up the $R_0$\ values to get estimated effective radii. 
The effective radii are shown in parentheses in Table 4, and the circles in
figures 1 - 10 are drawn with these radii.

\section{Reddening, Distance and Age}
We constructed colour magnitude diagrams (CMDs) using stars located 
within the effective cluster
radii. Figures 12-21 show the CMDs of both the clusters and the 
surrounding field regions,
consisting of stars located more than 3.5 arcmin from the cluster centers.  
Since these clusters
are sparse and embedded in rich galactic fields, our conservative 
choice of the cluster radii maximizes the contrast between the cluster and the
field population.

We compared the cluster CMDs with the 
Girardi et al.(2000) isochrones to estimate the cluster
reddenings, distances and ages. Without independent information about a 
cluster's reddening and distance, it is not possible to make an unambiguous
determination of the appropriate age and metallicity isochrone to fit to the
cluster.  Instead, we have made a visual determination of the most likely 
isochrone, or
range of isochrones, that appears to match the cluster.  The estimated 
isochrone parameters
are tabulated in Table 4, and the best fit of the selected isochrones
to the individual cluster CMDs (V vs (B$-$V)) are shown 
in figure 22 and figure 23 (V vs (V$-$I)). The  parameters estimated
by Tadross (2008) for the common clusters are tablulated in Table 5,
for comparison.

\begin{figure} 
\epsfxsize=9truecm
\epsffile{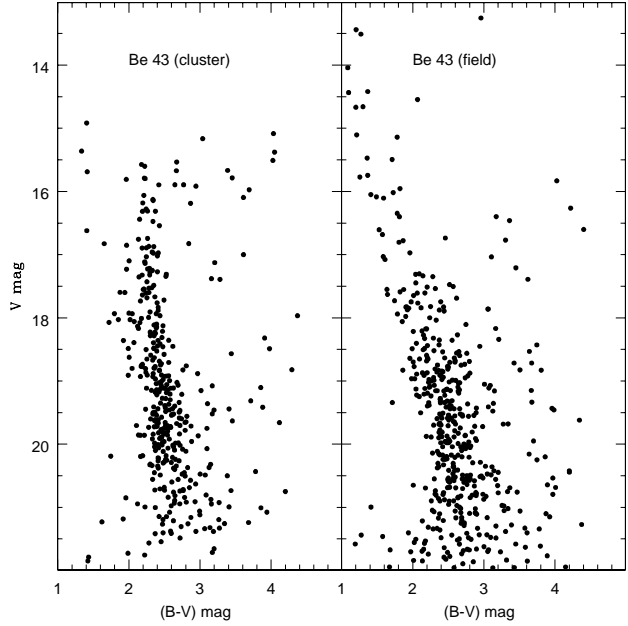} 
\caption{The V vs (B$-$V) CMDs of the cluster Berkeley 43 (left) and field 
region (right). We assumed that stars located
beyond a radius of 3.5 arcmin are field population.
}
\end{figure} 

\begin{figure} 
\epsfxsize=9truecm
\epsffile{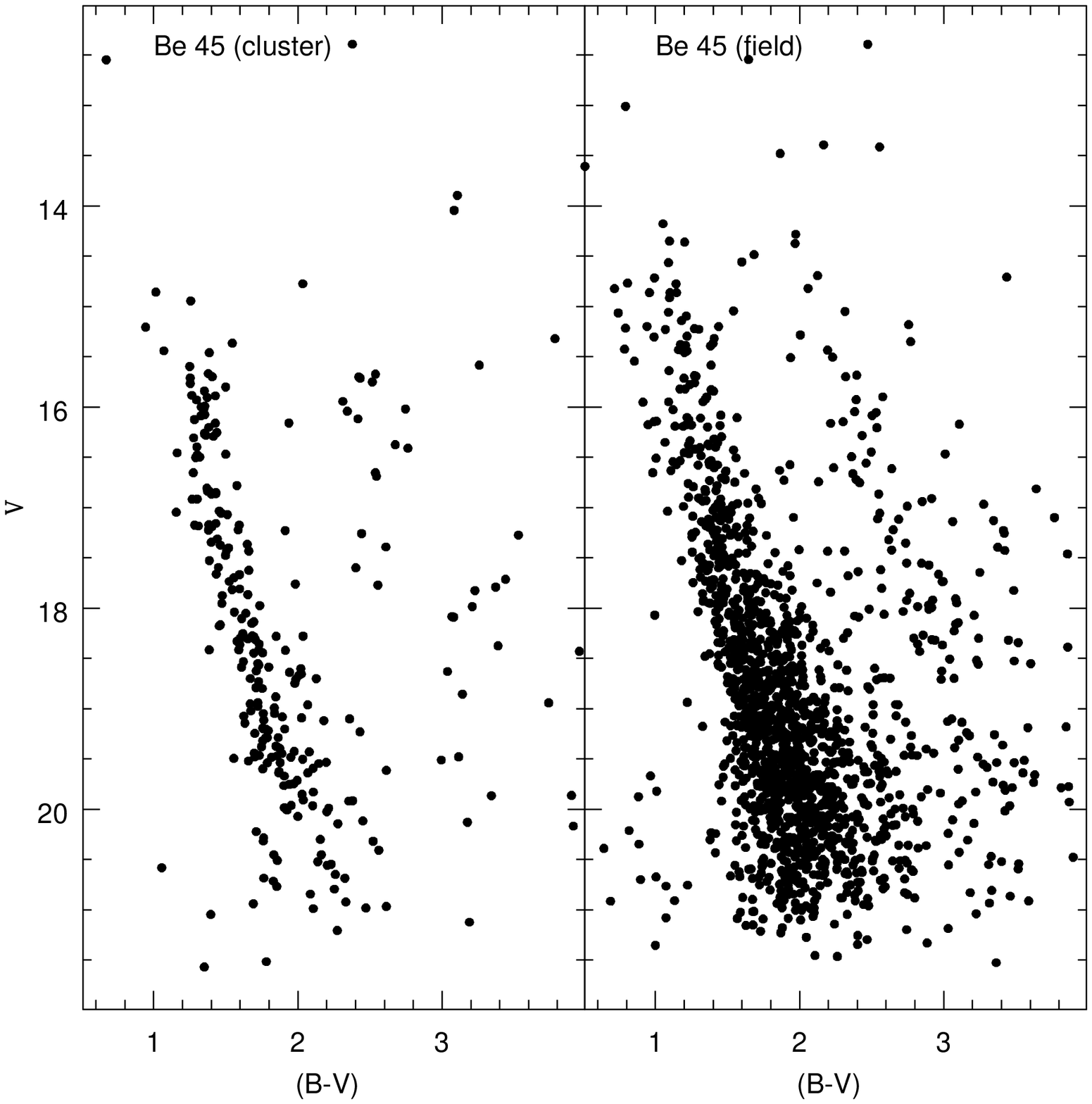} 
\caption{The V vs (B$-$V) CMDs of the cluster Berkeley 45 (left) and field 
region (right). We assumed that stars located
beyond a radius of 3.5 arcmin are field population.
}
\end{figure} 

\begin{figure} 
\epsfxsize=9truecm
\epsffile{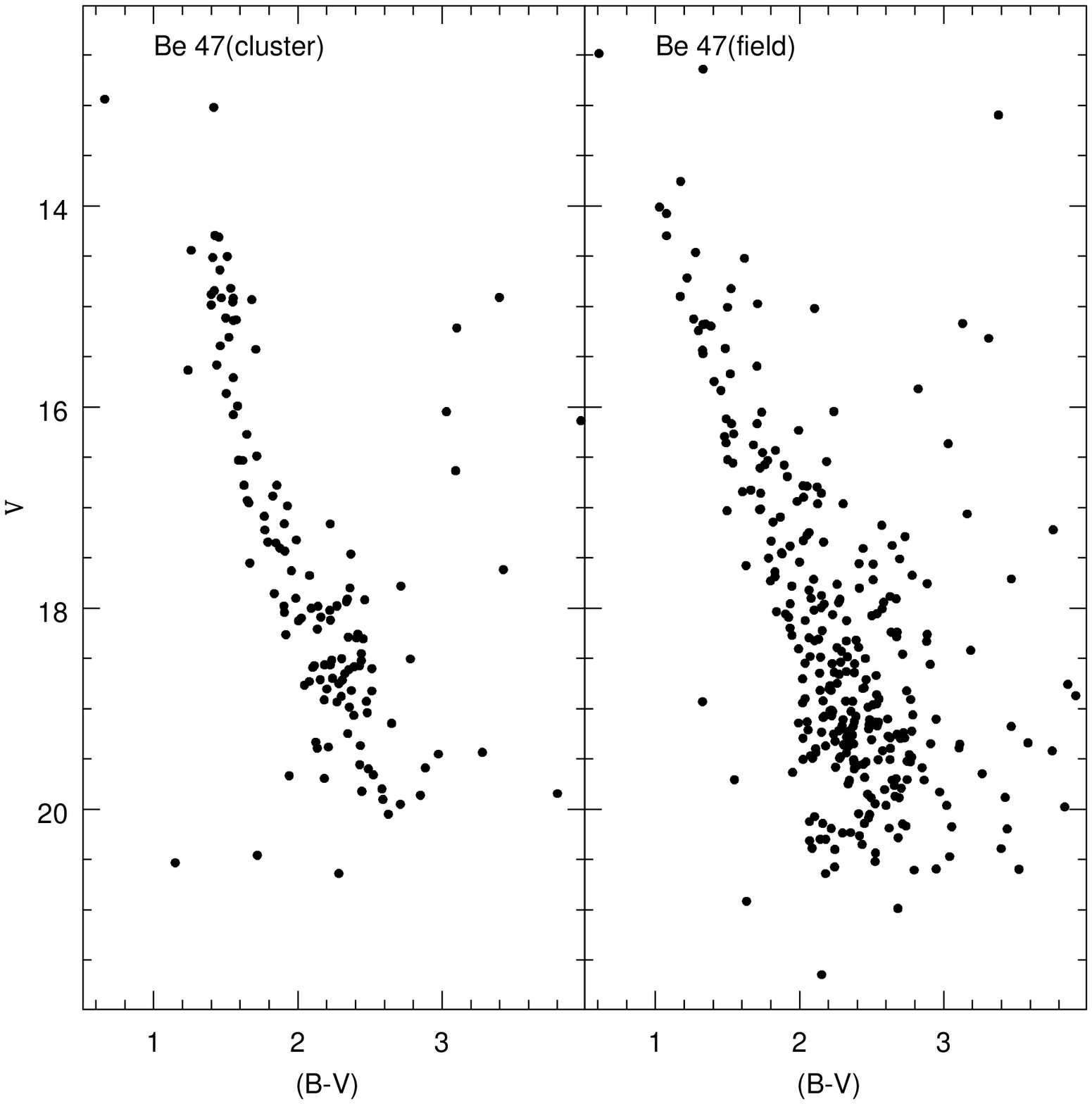} 
\caption{The V vs (B$-$V) CMDs of the cluster Berkeley 47 (left) and field 
region (right). We assumed that stars located
beyond a radius of 3.5 arcmin are field population.
}
\end{figure} 

\begin{figure} 
\epsfxsize=9truecm
\epsffile{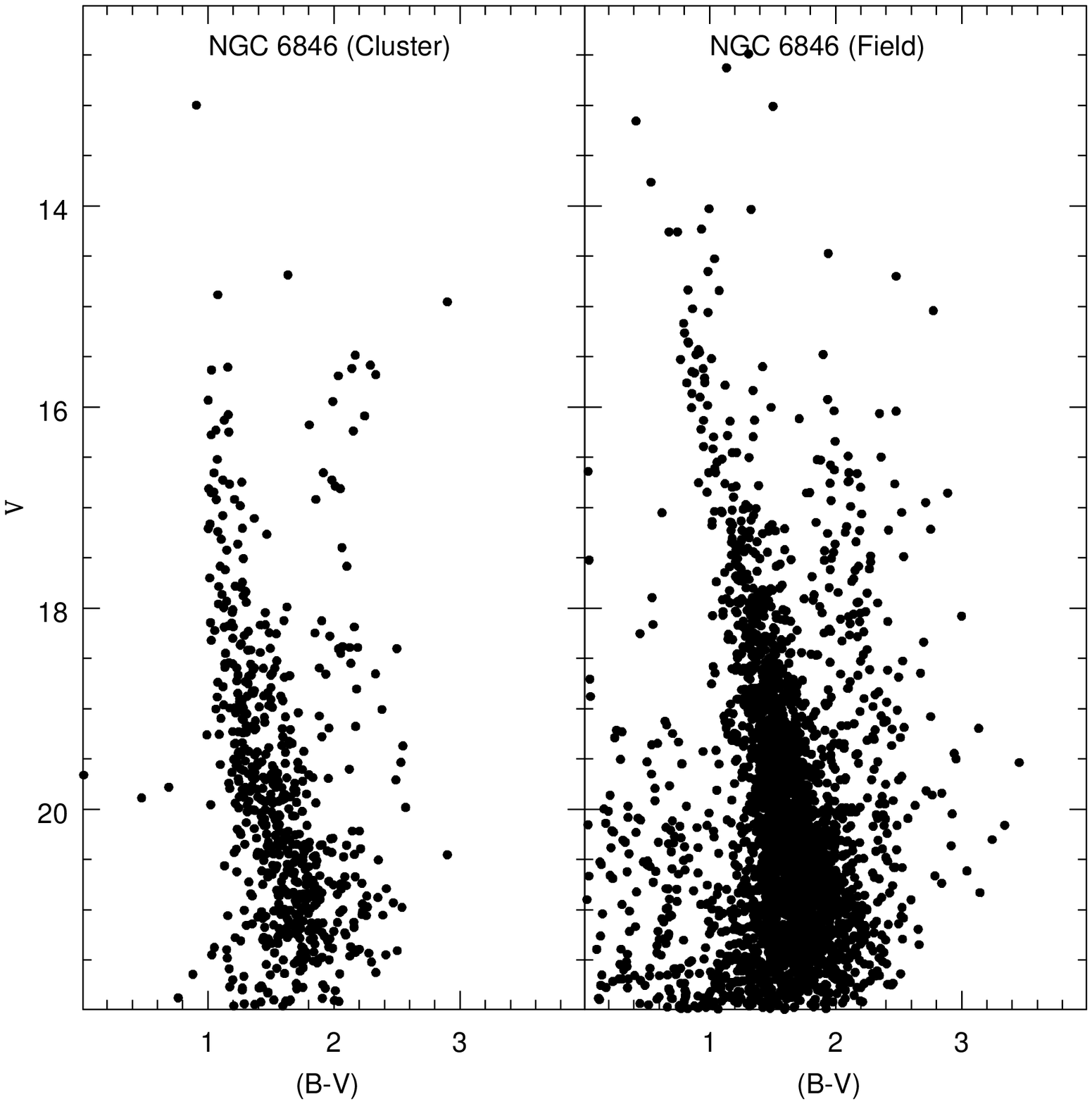} 
\caption{The V vs (B$-$V) CMDs of the cluster NGC 6846 (left) 
and field region (right). We assumed that stars located
beyond a radius of 3.5 arcmin are field population.
}
\end{figure} 

\begin{figure} 
\epsfxsize=9truecm
\epsffile{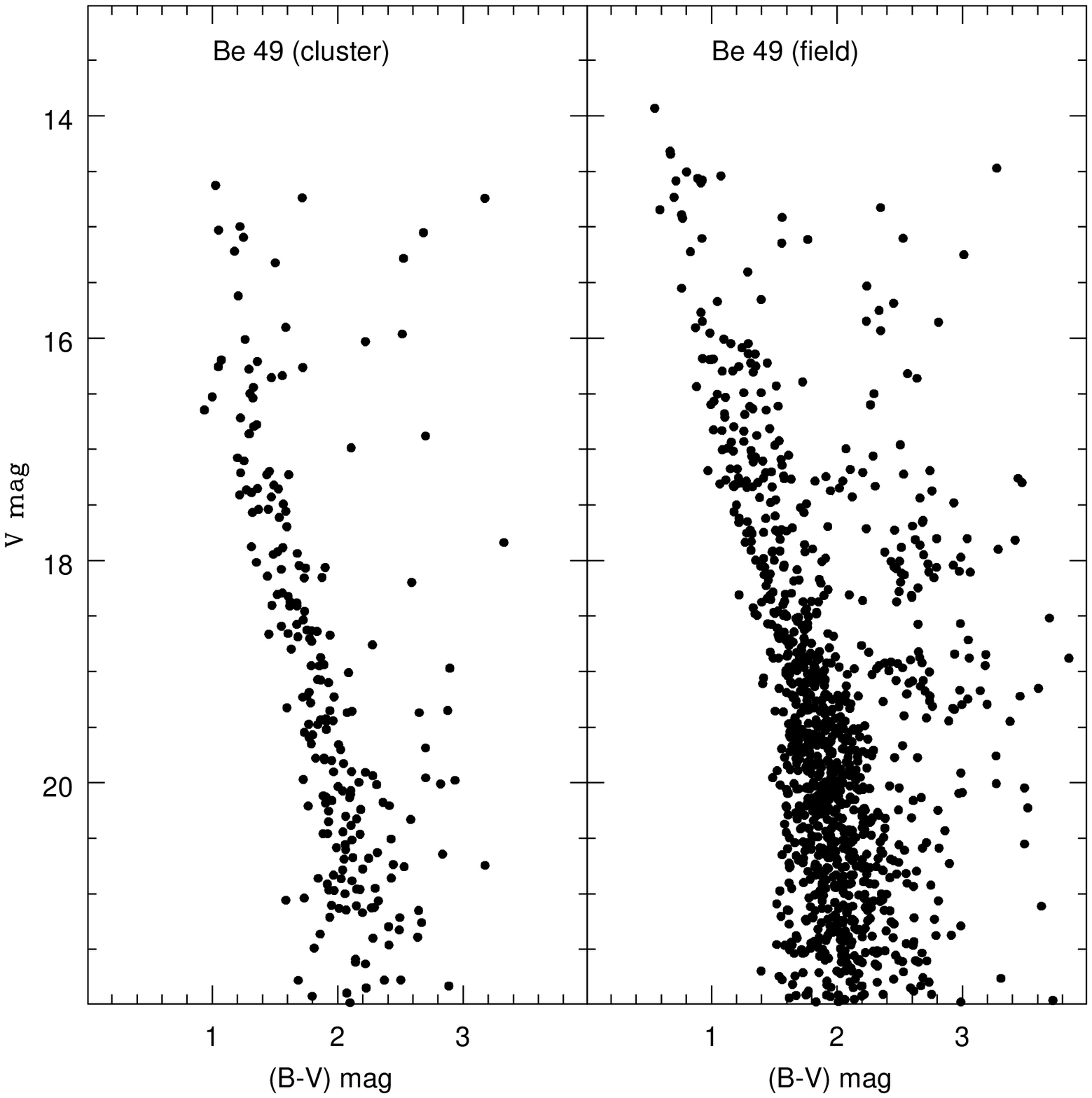} 
\caption{The V vs (B$-$V) CMDs of the cluster Berkeley 49 (left) and 
field region (right). We assumed that stars located
beyond a radius of 3.5 arcmin are field population.
}
\end{figure} 
\begin{figure} 
\epsfxsize=9truecm
\epsffile{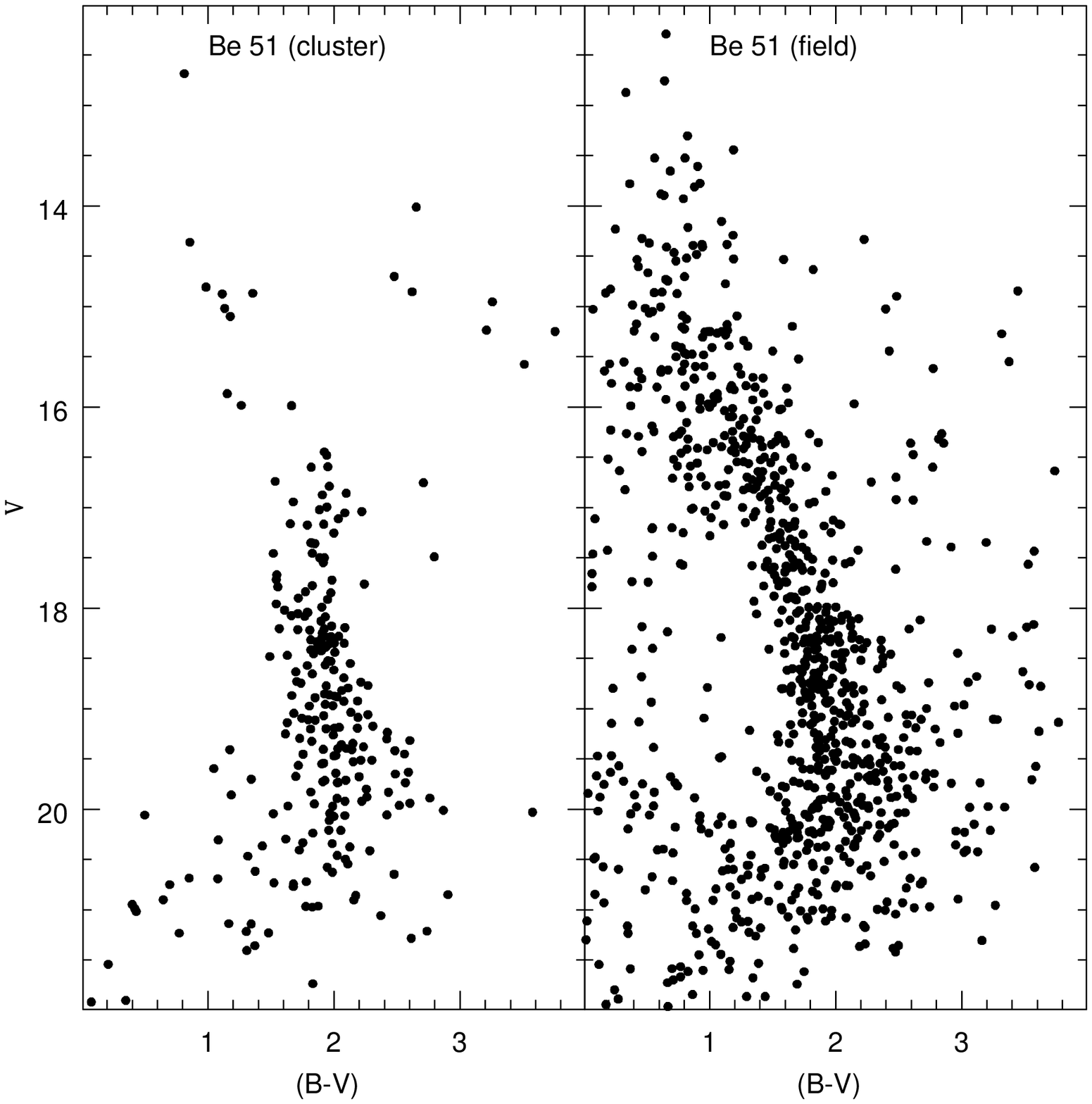} 
\caption{The V vs (B$-$V) CMDs of the cluster Berkeley 51 (left) and field 
region (right). We assumed that stars located
beyond a radius of 3.5 arcmin are field population.
}
\end{figure} 
\begin{figure} 
\epsfxsize=9truecm
\epsffile{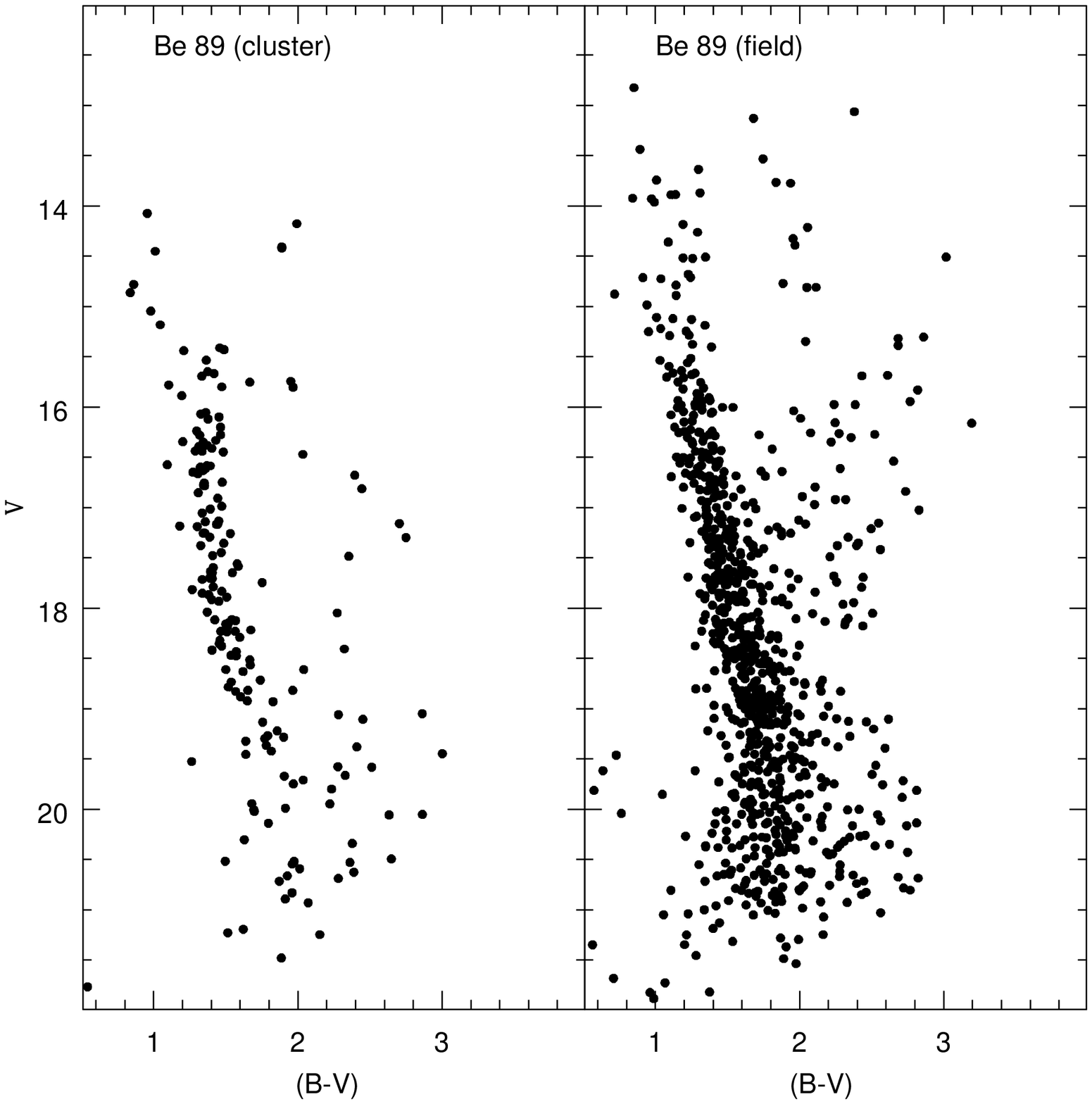} 
\caption{The V vs (B$-$V) CMDs of the cluster Berkeley 89 (left) and field 
region (right). We assumed that stars located
beyond a radius of 3.5 arcmin are field population.
}
\end{figure} 
\begin{figure} 
\epsfxsize=9truecm
\epsffile{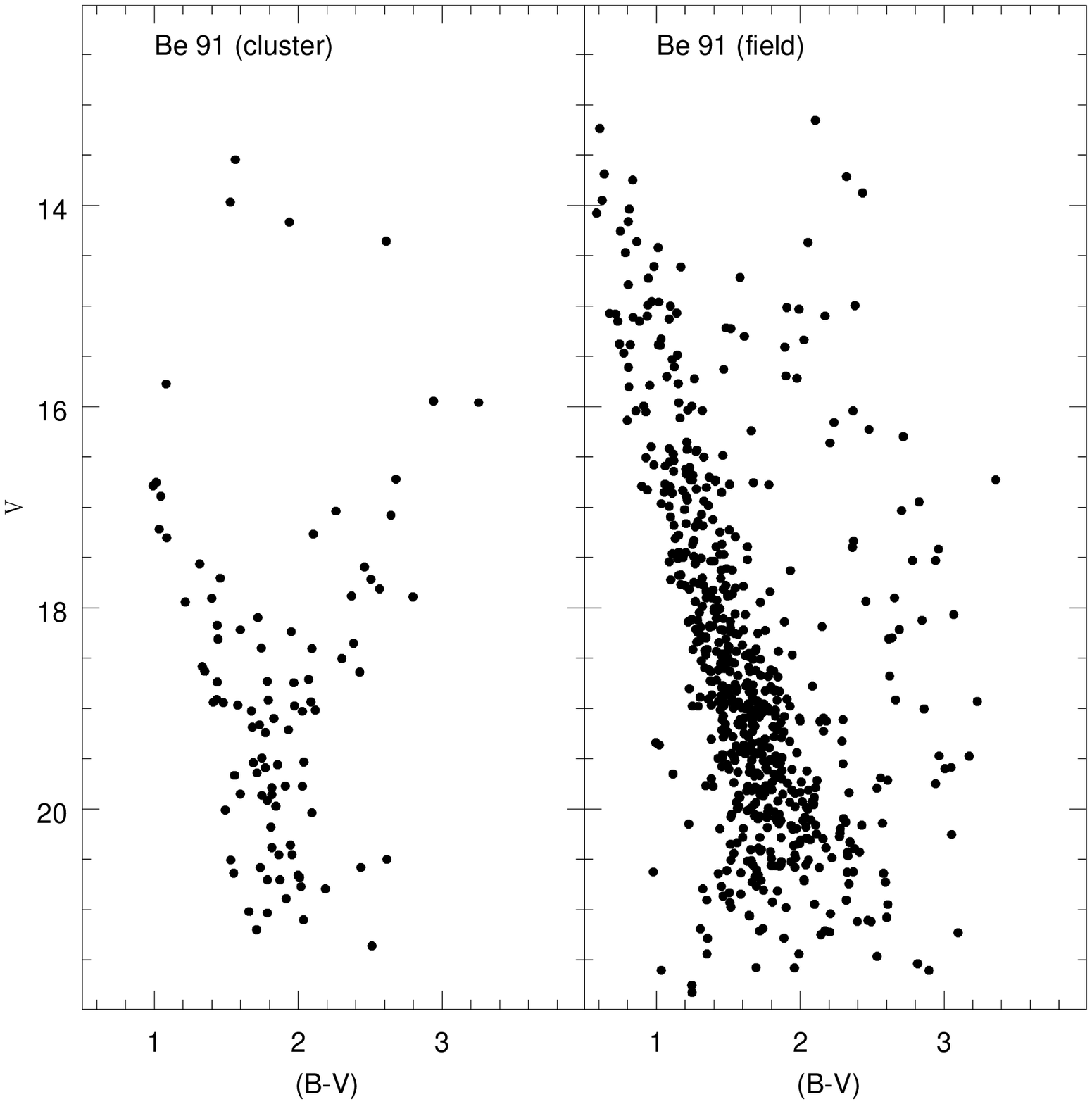} 
\caption{The V vs (B$-$V) CMDs of the cluster Berkeley 91 (left) and field 
region (right). We assumed that stars located
beyond a radius of 3.5 arcmin are field population.
}
\end{figure} 

\begin{figure} 
\epsfxsize=9truecm
\epsffile{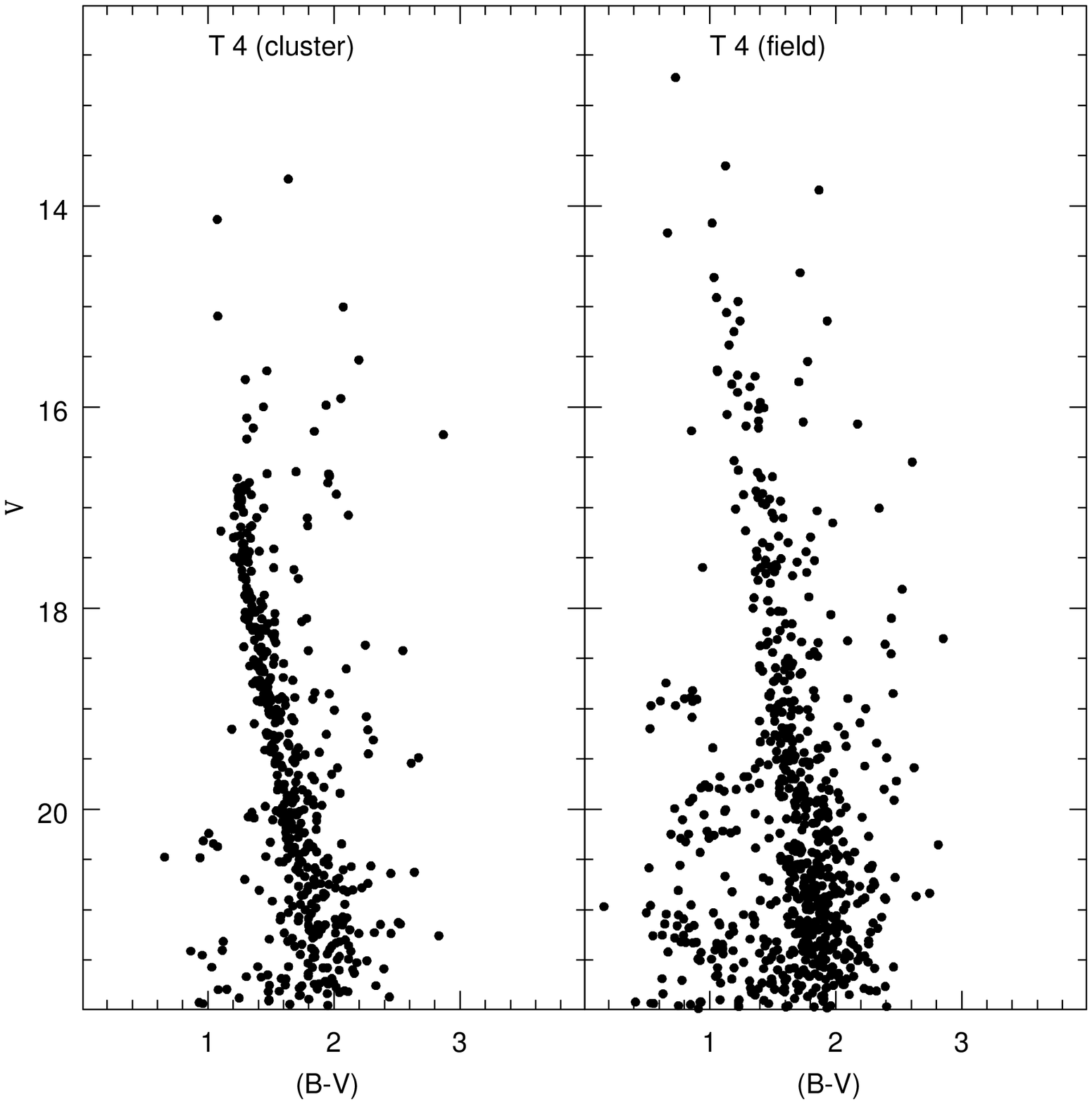} 
\caption{The V vs (B$-$V) CMDs of the cluster Tombaugh 4 (left) and field 
region (right). We assumed that stars located
beyond a radius of 3.5 arcmin are field population
}
\end{figure} 
\begin{figure} 
\epsfxsize=9truecm
\epsffile{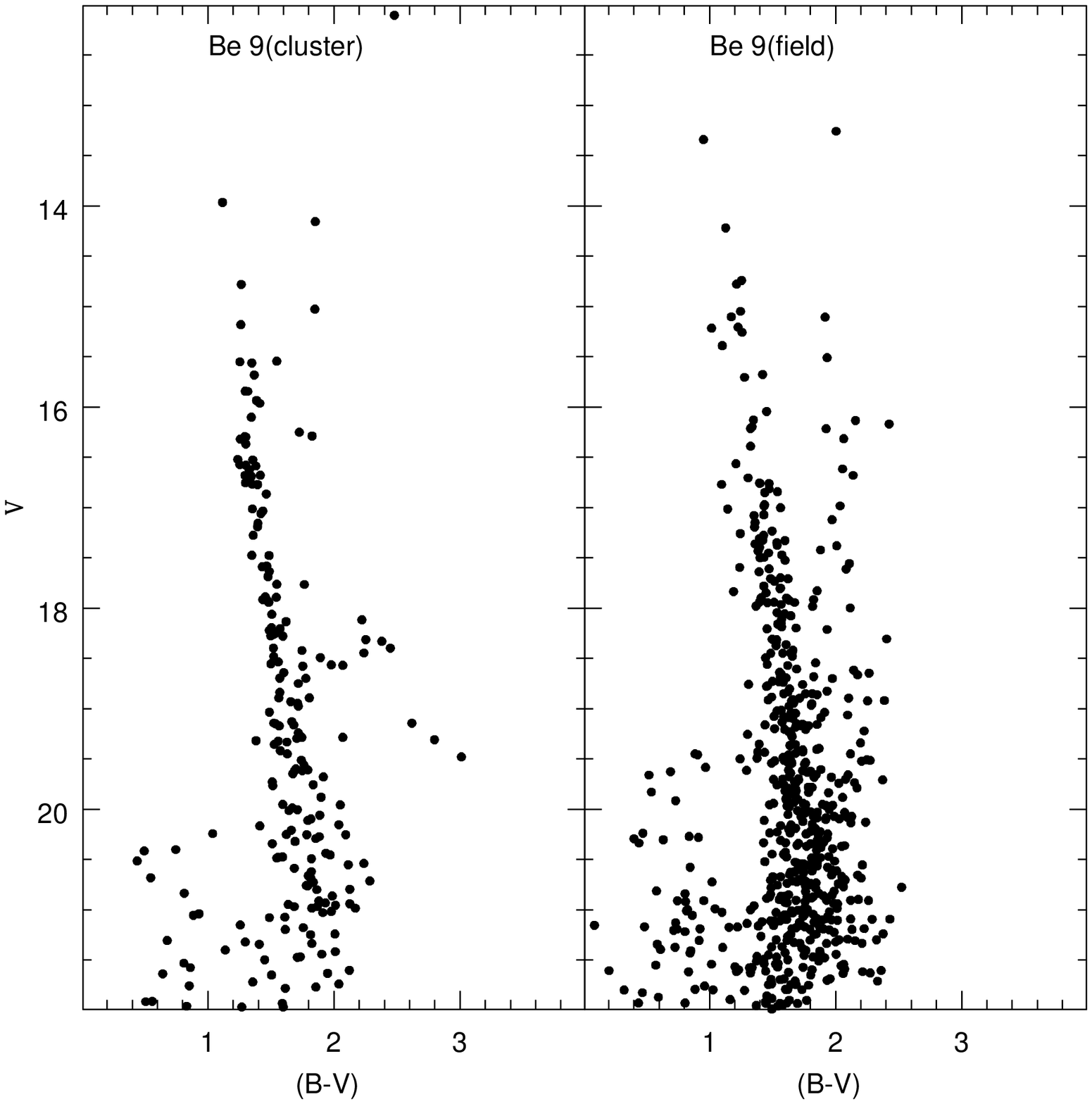} 
\caption{The V vs (B$-$V) CMDs of the cluster Berkeley 9 (left) and field 
region (right). We assumed that stars located
beyond a radius of 3.5 arcmin are field population
}
\end{figure} 

{\bf Berkeley 43}: The cluster is located near the north side of the observed 
region (Figure 1).
The stars in the cluster are distributed in two concentrations in the 
north-south direction.
A deficiency of stars on the south-eastern
side of the cluster is probably due to the presence of greater extinction, so
the cluster may be located near a dark cloud.
This is supported by the high reddening we estimate towards this cluster 
(Table 4). 
The colour-magnitude
diagrams (CMD) of likely cluster stars and stars located 
beyond a radius of
3.5 arcmin (assumed to be field stars) are shown in figure 12.
The cluster
sequence is  distinctly identifiable, when compared to the field.
The isochrone compared to the cluster (figures 22 \& 23) suggest 
a relatively nearby ($\sim$ 1 kpc) young cluster (log(age) = 8.4-8.5), 
very
close to the mid galactic plane. Hasegawa et al. (2008) estimated a distance 
of 1500 pc,
E(V$-$I) of 1.96 mag and an age of 1.4 Gyr. We find a much younger 
age for this cluster.
The cluster properties estimated by Tadross (2008), using 2MASS data 
match broadly with
our estimates. 

{\bf Berkeley 45}: 
The cluster is located in a fairly rich galactic field (figure 2),
visible as a
marginal enhancement in the density from the field; it could be a chance
asterism.
Nevertheless, the presence of the cluster is clearly seen in the
radial density profile (figure 11). A
decrease in the star density toward the north-west suggests variable
extinction in the field. This is a
sparse cluster and the cluster and field
CMDs are not very different, so it is difficult to 
identify the MS in the CMD (figure 13).   
The only clear difference is the narrow 
MS with a distinct cut-off/turn-off
at the bright end. Therefore, we consider this as a true cluster.  The 
CMD of the
cluster region is well-matched by the isochrones suggesting a young 
(log(age)=8.5-8.6),
cluster, located at an approximate distance of 2350 pc with considerable reddening 
(E(B$-$V) = 1.4 mag).
In an earlier study,
Tadross (2008) estimated a distance of 2.3 kpc and an age of 600 Myr, 
using 2MASS data. Our estimate of the distance is within errors, but we
estimate a marginally younger age for the cluster.

{\bf Berkeley 47}: This cluster is located close to Berkeley 45, in the 
first quadrant. It is a rather poor cluster, but it
stands out from the somewhat sparse field (figure 3).
Although the radial density profile clearly shows the presence of the 
cluster, it may also be an asterism.
There is a decrease in the number of field
stars on the north-west and south-east sides of the cluster, suggesting that 
the cluster may be located between two dark clouds. 
The cluster CMD has a well defined MS, when compared to the field 
CMD (figure 14). There are no giants present in the CMD
and the cluster parameters are estimated based on the MS and its turn-off.
The isochrone fit to the cluster CMD shows a poor match to the
lower MS, possibly because of field star contamination.
Another possibility is
differential reddening, which could be present across the face of the cluster.
The fainter stars are located away from the cluster center (figure 3)
and hence may be more reddened.  Since this is a poor cluster, it is difficult  
to attribute the poor fit to any model limitations or to arrive at any
definite possibility for the observed deviation.

 We find Berkeley 47 to be a moderately
young (log(age) = 8.6-8.7) cluster, located at $\sim$ 1 kpc with 
large reddening (E(B$-$V) = 1.5 mag) .

{\bf NGC 6846} This is a
compact and distant cluster embedded in a rich
galactic field, including a number of bright foreground stars.
The region towards the south of the cluster has relatively lower 
field star density, indicating variable interstellar extinction. 
The cluster and the 
field CMDs (figure 15) are rather different from one another.  There are a 
few red giants in the cluster
CMD, some of which may be members. 
The most likely isochrone fit to the cluster suggests
that it is moderately young (log(age) = 8.6-8.7), 
located about 5.1 kpc away,
with substantial reddening (E(B$-$V) = 1.05 mag). This cluster is 
located well above the galactic plane (Z $\sim$ 170 pc).
This cluster is most distant cluster in the first quadrant, more distant
than IC 1311, an older cluster 
located at about 4.0 kpc (Delgado et al. 1994).
The stars in NGC 6846 are potential targets to study
the abundances of the Galactic disk in these regions.

{\bf Berkeley 49} This cluster is also in a rich
galactic field and does not stand out from the field, especially 
because of its
clumpy nature.  
It can be seen from figure 5, that there is a gap with fewer stars located 
towards the north-west of the cluster. Like several of the other
clusters, this one is located in a region of variable extinction.
This is supported
by the high reddening estimated for the cluster (E(B$-$V) = 1.35 mag). 
The radial density profile
suggests a dense core and a very small radius for the cluster. 
The cluster CMD shows a narrow MS when compared
to the field CMD (figure 16), even though the field and the cluster CMDs 
do not look very different. There is a possibility that this cluster is
a chance grouping of stars.
The isochrone fits to the
cluster CMD suggests that Berkeley 49 is a young cluster 
(log(age) = 8.35- 8.5), located about 2.3 kpc away.
In a previous study,
Tadross (2008) estimated a distance of 2.0 kpc and an age of 160 Myr, using 
2MASS data,
which are similar to the present estimates.

{\bf Berkeley 51} This cluster is located close to Berkeley 49 
in the galactic mid plane.
It is rich and compact, but like several others in this sample it is
located in a clumpy galactic field. Adjoining regions
towards the north-west and south show fewer field stars, indicating
variable extinction.
The cluster is easily identified in the field due to its high density. 
The radial density
profile shows that the cluster has a dense core and and a small radius.
The cluster and the field CMDs are shown in figure 17. 
The cluster CMD is very
different from the field CMD. The short and faint MS shows an older
cluster embedded in a galactic field rich in younger stars.
The best isochrone comparison to
the cluster CMD confirms that Berkeley 51 is probably
a moderately old cluster (log(age) = 9.0 - 9.05), 
located at a distance of 1.3 kpc, with
large reddening (E(B$-$V) = 1.6 mag). In a previous study, 
Tadross (2008) estimated a distance of 3.2 kpc and an age of 150 Myr, 
using 2MASS data.
This does not agree with the present estimations. 
Since it is a 
cluster of faint stars located in 
a rich and bright galactic field, it is very likely that the parameters 
estimated by Tadross (2008) refer to the
majority of field stars and not the faint cluster members.

{\bf Berkeley 89} This cluster is located at a somewhat 
higher galactic latitude than the other clusters.
It is relatively poor (figure 7) located in a rich galactic field, and only
marginally distinct from the field. This cluster may also be an asterism. 
Nevertheless, a cluster sequence is clearly
visible, when compared to the field CMD (figure 18), and
the radial density profile shows a cluster although without
a dense core, gradually merging with the field 
population. 
The best-matching isochrone indicates that the cluster is similar in age
to Berkeley 51 with log(age) = 9.0- 9.05.
Berkeley 89 is located at a distance of 2.0 kpc, with a reddening
of E(B$-$V) = 1.05 mag. 
In his previous study Tadross (2008) estimated a 
distance of 3.0 kpc and an age of $\sim$ 1 Gyr, using 2MASS data.
The cluster is dynamically relaxed and mass
segregation is likely to be present. This leads to the spatial separation of low mass
stars in the outer regions and high mass stars near the cluster center.
Since we have considered a radius which is not much larger than the
core radius, we probably do not sample enough low mass stars to populate the
lower MS.

{\bf Berkeley 91} 
The cluster is a relatively poor cluster 
with relatively faint stars in a rich field. 
The radial density profile shows that this 
is a sparse but small cluster.  It may not be a real cluster.
In order to  identify the cluster sequence clearly, we used the core 
radius of the cluster to identify
cluster stars. 
The cluster CMD shows a well defined MS, a short sub-giant 
branch and a red giant branch (figure 19).
If it is real, it is clearly the oldest of the sample 
studied here. The cluster sequence is also
very different than the field CMD.  The cluster CMD is well-matched to
old cluster isochrones (log(age) = 9.15 - 9.35), yielding a distance of 
4.1 kpc.  

{\bf Tombaugh 4}
This is a fairly rich, compact cluster with several bright stars (figure 9). 
The radial density profile shows that the cluster
has a dense core. The field is  clumpy with an uneven
distribution of stars.
The cluster CMD has a narrow and well defined MS, when compared
to the  field CMD (figure 20). The cluster MS has a sharp turn-off. 
and a few red 
giants may be
cluster members.
The isochrone fits suggest that Tombaugh 4 is a moderately young
cluster (log(age) = 8.6-8.7), located at a distance of 3.85 kpc. 

{\bf Berkeley 9}
This is an irregular cluster located in a somewhat sparse
galactic field. It appears only as a mild density enhancement in the field. 
The cluster MS is found to be well 
defined (figure 21), even though there is not much of difference between the
cluster and the field CMD. The isochrone fit suggests that the cluster is 
young, (log(age) $\sim$ 7.8 - 8.0) and 
located at a distance of 2.3 kpc. This is the only cluster in our 
sample, about 100 Myr old and located below the galactic plane 
(Z = $-$113 pc).
Maciejewski \& Niedzielski (2007) studied Be 9 and found the cluster
to be very old (log (age) = 9.6), with E(B$-$V) = 0.76 and DM = 12.03. They identify a 
much fainter turn-off for the cluster.  We find a good number
of bright MS stars, above the turn-off identified by Maciejewski and
Niedzielski and hence we estimate 
a much younger age for this cluster. This is also supported by the V vs (V$-$I) CMD.

\renewcommand{\thetable}{4}
\begin{table*}
\caption{Fundamental parameters estimated for the clusters. Along with the 
half-power radius,
the effective 
radius used to estimate the cluster parameters in given in the parenthesis.
The error in the reddening estimate is 0.02 mag.
Errors in distances is about 10\%.} 
\begin{tabular}{rrrrrrrr}
\hline
Name &  Radius & E(B$-$V) & E(V$-$I) & DM & Distance & Log (Age) & Z \\
 & arcmin & (mag) & (mag) & (mag) & (pc) &  & (pc) \\
\hline
Berkeley 43   & 1.9(2.5) & 2.3 & 2.5   & 17.2 & 1030 & 8.45 - 8.5 & $\sim$ 0 \\
Berkeley 45   & 1.1(1.5) & 1.4 & 1.5   & 16.2 & 2350 & 8.5 - 8.6  &  47 \\
Berkeley 47   & 1.1(2.0) & 1.5 & 1.7   & 14.8 & 1070 & 8.65 - 8.7  & $\sim$ 0 \\
NGC 6846      & 1.5(1.5) & 1.05 & 1.35 & 16.5 & 4450 & 8.6 - 8.7  & 150\\
Berkeley 49   & 0.7(1.5) & 1.35 & 1.55 & 16.0 & 2300 & 8.35 - 8.5 &  103\\
Berkeley 51   & 0.6(1.0) & 1.6 & 1.6   & 15.5 & 1300 & 9.0 - 9.05 & $\sim$ 0\\
Berkeley 89   & 0.8(1.0) & 1.05 & 1.6  & 14.8 & 2040 & 9.0 - 9.05 & 171 \\
Berkeley 91   & 1.0(1.0)& 1.2 & 0.6    & 16.8 & 4100 & 9.15 - 9.35 & $\sim$ 0  \\
Tombaugh 4    & 1.3(2.5) & 1.25 & 1.6  & 16.8 & 3850 & 8.6 - 8.7  & 72 \\
Berkeley 9    & 1.0(2.0) & 1.45 & 1.65 & 16.3 & 2300 & 7.8 - 8.0  & $-$113\\
\hline
\end{tabular}
\end{table*}

\renewcommand{\thetable}{5}
\begin{table*}
\caption{Cluster parameters as estimated by Tadross (2008), using 2MASS data.
} 
\begin{tabular}{rrrrr}
\hline
Name &   Radius & E(B$-$V) &  Distance & Log (Age)  \\
     & arcmin       & (mag)    &   (pc)     &           \\
\hline
Berkeley 43 & 4.5 &  1.52 & 1355$\pm$60 & 8.6  \\
Berkeley 45 & 3.5 &  0.82 & 2300$\pm$105 & 8.77 \\
Berkeley 47 & 2.0 &  1.06 & 1420$\pm$65  &  8.2 \\
Berkeley 49 & 2.4 &  1.57 & 2035$\pm$110 & 8.2 \\
Berkeley 51 & 1.5 &  1.66 & 3200$\pm$145 & 8.17 \\
Berkeley 89 & 2.5 &  1.03 & 3005$\pm$135 & 8.9 \\
Berkeley 91 & 1.7 & 1.00 &  2400$\pm$110 & 8.7\\
\hline
\end{tabular}
\end{table*}


%
\begin{figure} 
\epsfxsize=15truecm
\epsffile{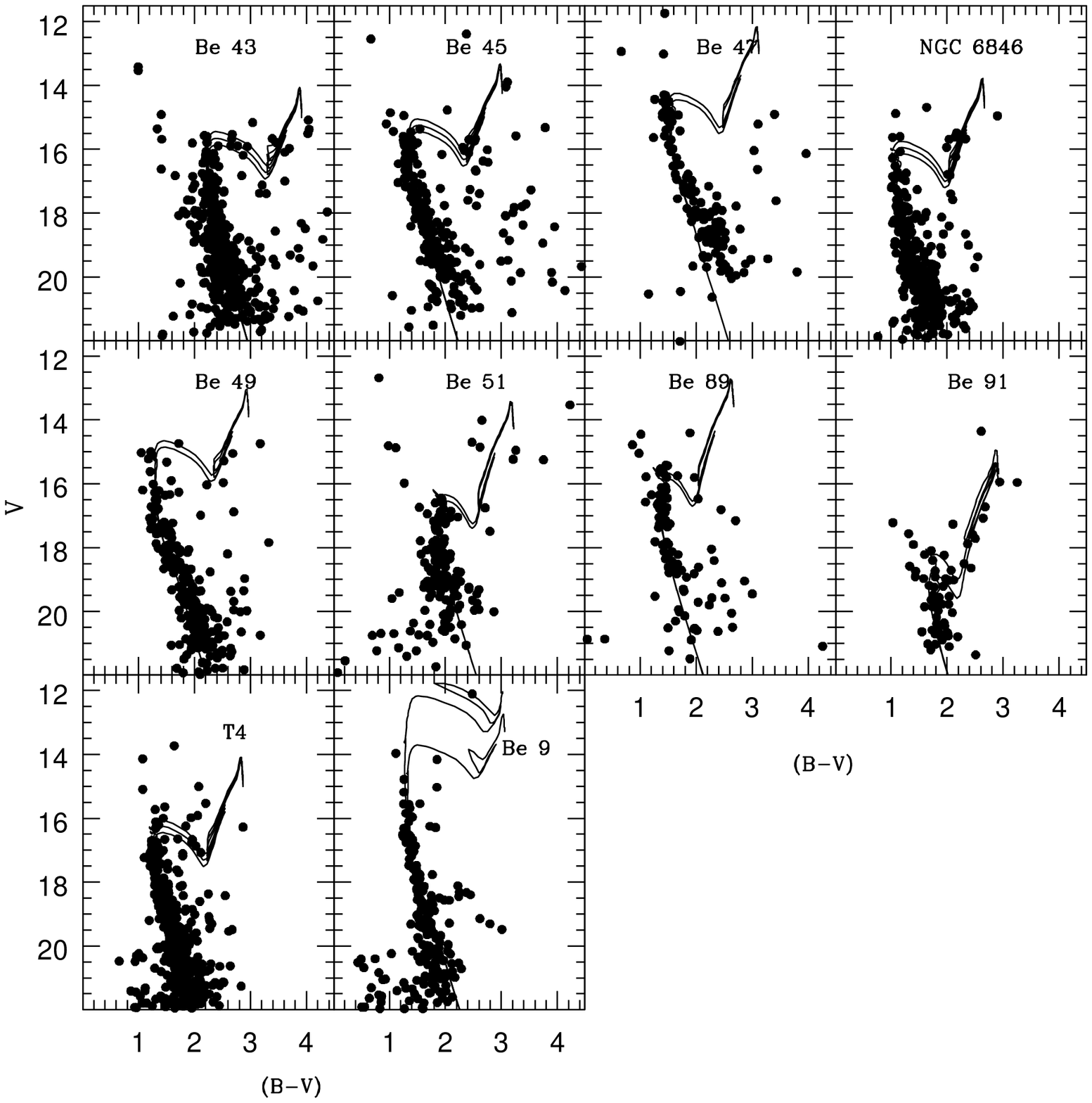} 
\caption{Isochrone fits to the  V vs (B$-$V) CMDs of ten clusters. We have used Girardi et al. (2000) isochrones.}
\end{figure} 
\begin{figure} 
\epsfxsize=15truecm
\epsffile{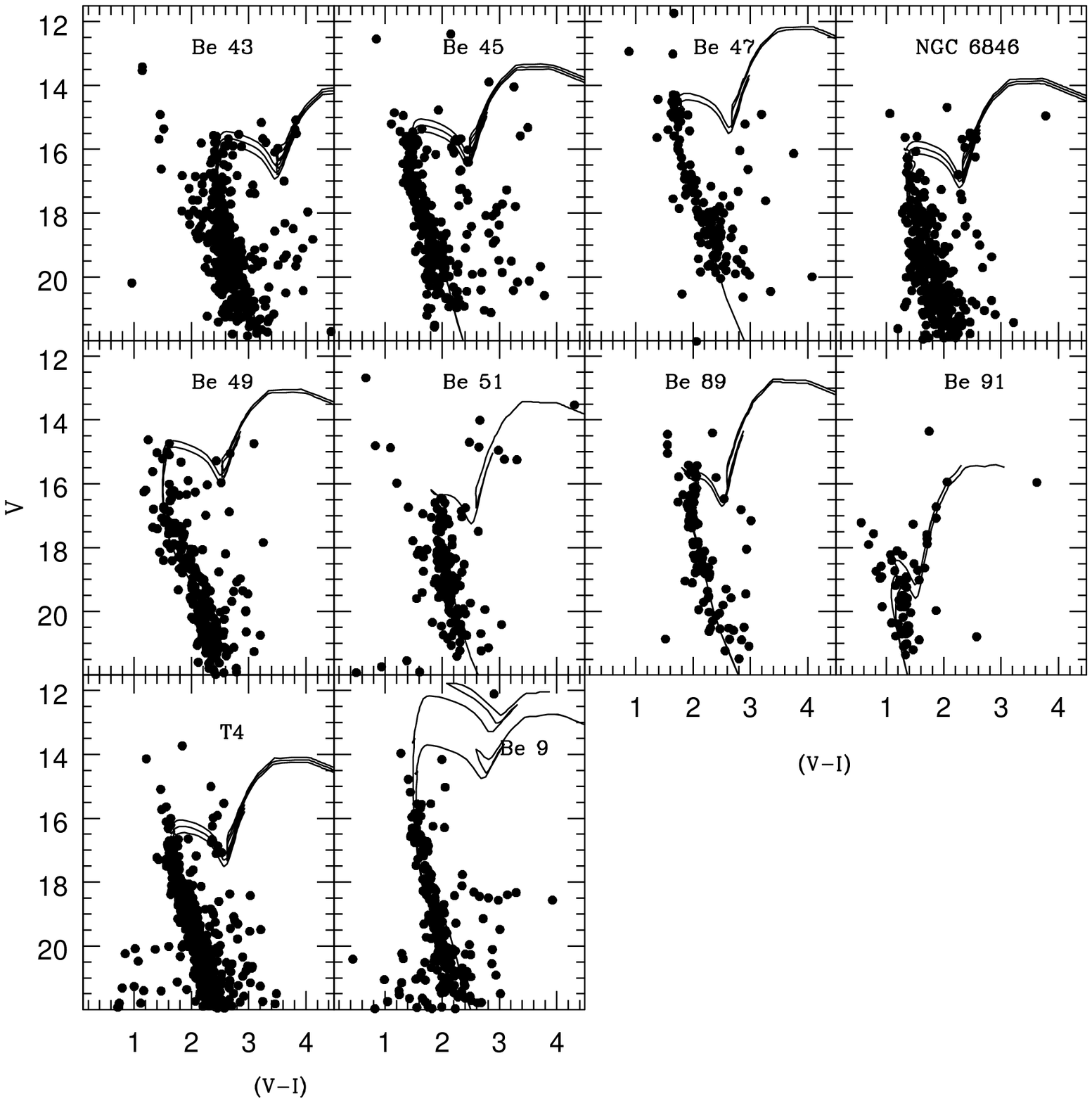} 
\caption{Isochrone fits to the  V vs (V$-$I) CMDs of ten clusters. We have used Girardi et al. (2000) isochrones.}
\end{figure} 
\begin{figure} 
\epsfxsize=9truecm
\epsffile{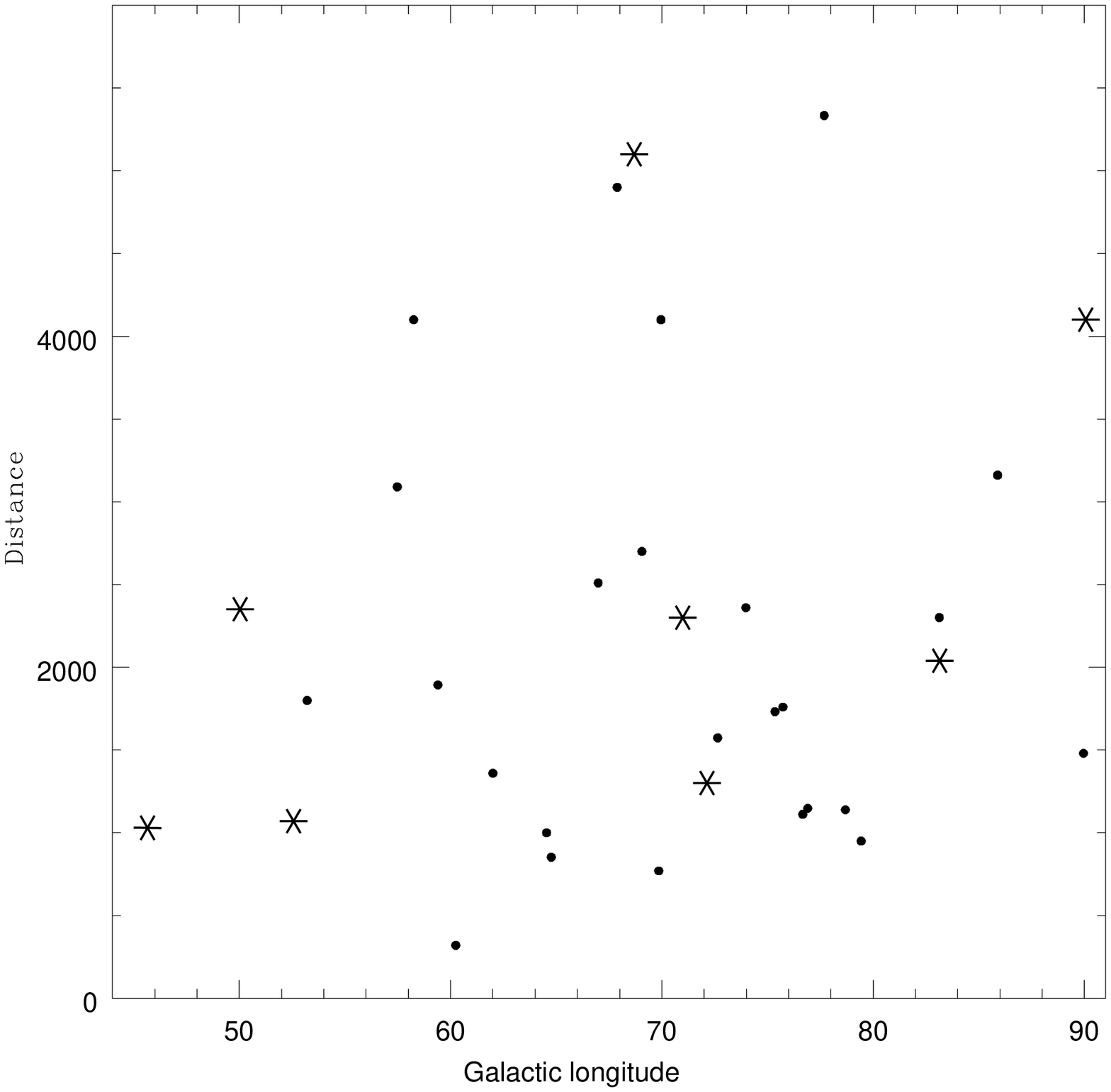} 
\caption{Clusters studied so far using UBV CCD photometry in the second galactic are shown as dots. Clusters
studied in this paper as shown as asterisks. 
}
\end{figure} 

\section{Discussion}
This study presents the basic parameters of 10 previously 
neglected open clusters.
The estimated parameters are summarised in Table 4.
Eight of the clusters studied here
are located in the first galactic quadrant, 5 of them beyond a 
distance of
2 kpc. There are 10 previously-studied clusters 
located between $ l = 45 - 90^o$, and farther than 2 kpc. This study 
adds another 5 (Dias et al. 2002).

A plot with the previously known clusters and the clusters studied 
here (asterisk) is
shown in figure 24. 
The only cluster studied previously using CCD photometry in the
range $l = 45-57^o$, is Berkeley 44 from our previous study, 
Carraro et al. (2006).
In the present study we have filled this gap with 3 more clusters. 
NGC 6846 is at at larger distance than 
IC 1311, (about 4 kpc away, Delgado, et al, 1994) previously the most
distant cluster known in this general direction.
Because of its distance and age, NGC 6846 is 
a potential candidate to study the properties of the young galactic disk in the
first quadrant.

Most of the clusters studied here are embedded in rich galactic fields 
and have large
reddening towards them (E(B$-$V) = 1.0 -- 2.3 mag). They are also all rather
sparsely populated clusters.
Thus these clusters are difficult to study, and in spite of the
appearance of the cluster CMDs and the radial density plots, there is a real
possibility that some of them are simply asterisms of random,
unrelated stars superposed on the galactic background. Without proper 
motions or spectroscopy, it is not possible to ascertain which of them might
be accidental stellar configurations. Future studies of these clusters
should note that some of them may not be real clusters. If some are found to be asterisms,
then their contribution to the understanding of the galactic structure, as discussed 
in this paper, are not valid.

Assuming they are all real our best estimates are that 6 of the clusters 
are likely to have ages less than 500 Myr and three more may be
as old as or older than 1 Gyr.  For comparison, Tadross (2008) estimated 
parameters of 7 clusters in common with this study 
using 2MASS data. The estimated parameters broadly match for 
Berkeley 43, Berkeley 45, Berkeley 49 
and Berkeley 89. The mismatch in the estimated parameters are probably a
consequence of the fact 
that they are located in rich galactic fields.
The parameters estimated by Hasegawa et al. (2008) for Berkeley 43 match
well, except that we find a much younger age for the cluster.

The clusters studied here could be used to trace the structure of the disk, 
especially
in the second half of the first quadrant.
An interesting result is that the 3 clusters at l=70-80 are above the
plane, as expected from our knowledge of the warp, which is found above 
the plane in the first
quadrant and below in the third 
both in optical (Momany et al. 2006) and in HI (Levine et al. 2006).

Tombaugh 4 is a moderately young cluster located in the second quadrant. 
Subramaniam \& Bhatt (2007) found an extension of the Perseus arm. This cluster
appears to be part of this extension. 

\section{Acknowledgment}
{A.S. acknowledges ESO support as
visiting astronomer to ESO, Santiago, where part of the work was conducted. 
Thanks to B. Mathew for observational support. G.C. acknowledges
ESO DGDF support during a visit to Boston University, where this paper
was prepared.}

\end{document}